\documentclass[11pt,a4paper]{article}

\usepackage[utf8]{inputenc}
\usepackage{amsmath}
\usepackage{amsfonts}
\usepackage{nameref, hyperref}
\usepackage{tikz}
\usetikzlibrary{shapes,arrows,calc,positioning}
\usepackage[ruled]{algorithm2e}
\usepackage[font=footnotesize,labelfont=bf]{caption}
\usepackage{multirow}
\usepackage{arxiv}

\RequirePackage{natbib}
\graphicspath{{fig/}{./fig/}}

\renewcommand{\v}[1]{\ensuremath{\mathbf{#1}}}
\newcommand{\gv}[1]{\ensuremath{\mbox{\boldmath$ #1 $}}}
\usepackage{soul}

\begin{document}

\title{Inference in  epidemiological agent-based models using ensemble-based data assimilation}

\author{
    Tadeo Javier Cocucci \\
    FaMAF, Universidad Nacional de Córdoba \\
    Córdoba, Argentina \\
    \texttt{tadeojcocucci@gmail.com} \\
    \And
    Manuel Pulido \\
    FaCENA, Universidad Nacional del Nordeste\\
    Corrientes, Argentina\\
    CONICET, Corrientes, Argentina\\
    \And
    Juan Aparicio \\
    Universidad Nacional de Salta \\
    Salta, Argentina \\
    INENCO, CONICET \\
    \And
    Juan Ruiz \\
    FCEN, Universidad de Buenos Aires \\
    Buenos Aires, Argentina \\
    CIMA, CONICET \\
    \And
    Ignacio Simoy \\
    Universidad Nacional de Salta \\
    Salta, Argentina \\
    INENCO, CONICET \\
    \And
    Santiago Rosa \\
    FaMAF, Universidad Nacional de Córdoba \\
    Córdoba, Argentina \\
}

\maketitle

\begin{abstract}
To represent the complex individual interactions in the dynamics of disease spread informed by data, the coupling of an epidemiological agent-based model with the ensemble Kalman filter is proposed. The statistical inference of the propagation of a disease by means of ensemble-based data assimilation systems has been studied in previous work. The models used are mostly compartmental models representing the mean field evolution through ordinary differential equations. These techniques allow to monitor the propagation of the infections from data and to estimate several parameters of epidemiological interest. However, there are many important features which are based on the individual interactions that cannot be represented in the mean field equations, such as  social network and bubbles, contact tracing, isolating individuals in risk, and social network-based distancing strategies. Agent-based models can describe contact networks at an individual level, including demographic attributes such as age, neighbourhood, household, workplaces, schools, entertainment places, among others. Nevertheless, these models have several unknown parameters which are thus difficult to estimate. In this work, we propose the use of ensemble-based data assimilation techniques to calibrate an agent-based model using daily epidemiological data. This raises the challenge of having to adapt the agent populations to incorporate the information provided by the coarse-grained data. To do this, two stochastic strategies to correct the model predictions are developed. The ensemble Kalman filter with perturbed observations is used for the joint estimation of the state and some key epidemiological parameters. We conduct experiments with an agent based-model designed for COVID-19 and assess the proposed methodology on synthetic data and on COVID-19 daily reports from Ciudad Autónoma de Buenos Aires, Argentina.

\end{abstract}
\section{Introduction}

Prediction models usually represent a system through a set of differential equations that govern the evolution of the system through continuous variables. Agent-based models (ABMs) rely on a different paradigm. They explicitly represent the characteristics and behaviour of interacting autonomous individuals --usually referred to as agents-- and use them to simulate  scenarios which serve as a modelization of complex systems \citep{Bonabeau2002}. Even simple interactions rules may lead to self-organization and emerging collective behaviour \citep{Helbing2012}. Therefore, ABMs follow a bottom-up approach in describing the dynamics of the system. They are useful to model the dynamics of epidemiological \citep{Vynnycky2010}, ecological \citep{Grimm2005}, economical \citep{Tesfatsion2006}, and social \citep{Epstein1996} systems. 

From a computational point of view, each agent consists of a collection of data attributes that describes agent's state. The behaviour and interaction of agents are governed via autonomous decisions and/or probabilistic rules which eventually modify the agent's status. ABMs can be interpreted from the perspective of object oriented programming (indeed, many implementations follow this paradigm) where each agent is an instance of a class with certain attributes. These attributes can be real numbers (e.g., space coordinates) or categorical variables (e.g., epidemiological state, social class), among other possible data types.  The possibility of describing complex behaviour through a potentially simple set of rules has fostered the popularity of ABMs in recent years. The computational cost is not negligible for systems with a large number of agents, however nowadays this is not a strong limitation due to the increase of computational power.

By describing the epidemiological state of individuals and their network of social contacts, epidemiological ABMs are suitable to represent quite realistically the evolution and spread of infections \citep{Roche2011}. The use of ABMs to represent infectious disease dynamics is promising because infections are indeed produced by contacts between people, and ABMs allow to model at this (micro)scale. In fact, it is quite straightforward to represent what is known about human interactions, through the interaction of agents. To model COVID-19 dynamics a number of ABMs have been developed which, among other features, include age structure and represent a social network that includes schools, houses, workplaces, etc. to provide realistic mixing (see, for example, \cite{Kerr2020,Flaxman2020,Simoy2021}).  Considering this social structure, the assessment of the effects of non-pharmaceutical interventions, such as confinement measures, closing of schools, social gathering limitations, can be captured and simulated with ABMs. A strong signature of the last decade has been the increase of devices (GPS, cameras, digitalized reports, commercial records) that can collect information at the human level. The potential of ABMs to model the complex interactions between individuals and to foster the use of anonymized individual-based information is huge. In this line, \cite{Aleta2020} use mobility and demographic data to construct the contact network and household distribution for a COVID-19 ABM which is used to assess the effect of non-pharmaceutical interventions.

One of the main limitations of ABMs is the need for setting  multiple simulation parameters. Recently, there were efforts to develop inference techniques to constrain ABM parameters through available observations. These are mainly focused on obtaining a proxy for the likelihood. In \cite{Hooten2020}, a variety of methods are proposed to calibrate ABMs, for instance, using Approximate Bayesian Computation alongside Markov chain Monte Carlo or approximating the likelihood using an emulator for the ABM.

The ensemble Kalman filter (EnKF) is a data assimilation (DA) technique suitable to conduct sequential Bayesian inference in noisy partially observed systems. Both, model predictions and observations, are assumed to have Gaussian uncertainties. The Gaussian assumption represents its main weakness and, at the same time, its main strength. Non-Gaussian uncertainties may result in a sub-optimal performance of the filter. On the other hand, a relatively small number of sample points --called ensemble members or particles-- may suffice for high dimensional state spaces. In particular, the correlation between variables can be well captured because of the Gaussian assumption. Observations of the system's state variables are often incomplete (not all state variables are observed) and indirect (observed variables are a function of the state variables). By considering the correlations between state variables, the EnKF can use the observations to improve the estimates of every state variable, even those which are not observed \citep{Carrassi2018}. Furthermore, unknown parameters of the model can be treated as unobserved variables and included in the state. Therefore, if there is enough correlation between the parameters and the observed variables, the EnKF is able to produce estimates of the unknown model parameters. This procedure, called state augmentation \citep{Annan2004, Ruiz2013}, is quite straightforward to implement.

Although the EnKF has  been originally developed for numerical weather prediction \citep{Evensen1994,Houtekamer1998}, its field of application has broadened over time. In particular, some works have applied the EnKF in epidemiological systems. \cite{Shaman2012, Shaman2013} used the ensemble adjustment Kalman filter (EAKF) to forecast the times at which the peaks for influenza outbreaks were reached. More recently, there are applications of the EnKF to infer COVID-19 transmission dynamics, \cite{Li2020} applied iterated filtering using the EAKF to estimate undocumented cases in China. Iterated filtering was earlier introduced by \cite{Ionides2006} based on particle filters with state augmentation, to provide maximum likelihood estimates of model parameters and used it to study cholera dynamics. \cite{Evensen2020} proposed the use of an ensemble Kalman smoother with multiple data assimilation (ESMDA) to estimate the parameters, mainly focused on the effective reproduction number, of a COVID-19 compartmental model. \cite{Ghostine2021} use an EnKF to study the impact of vaccination for COVID-19 using data from Saudi Arabia. All of these contributions use compartmental epidemiological models represented through differential equations. 

The application of ensemble-based DA concepts to combine ABMs with data  raises several challenges. A pioneer work in this direction combining the ensemble Kalman filter with an ABM  was conducted by \cite{Ward2016}. In that work, an ABM, combined with an ensemble Kalman filter to assimilate data from footfall cameras, was used to study pedestrian behaviour. They obtained satisfactory results, but the authors also note the challenges presented by parameter sensitivity and the need for parallelization when using large models. In ABMs, the model state is defined at a micro-scale as the set of current values of the attributes for each agent. The model provides the evolution of the attributes of each agent which are modified in time as a result of the interactions between them. For decision making, the individual state of each agent is not necessarily of interest and what is relevant are aggregated quantities and/or an anonymized group of individuals representing a certain class of attributes. These variables summarize information on the population as a whole. Even when the hidden state is at the micro-scale, and the inference's goal is to represent this microscopic state as closely as possible, observations are usually of the macroscopic scale so that the microscopic state is not well constrained by observations. We propose to conduct DA in the space of these aggregated variables. The mapping of the micro into the macro state is straightforward by aggregating the variables of interest for assimilation. But this is not the case for the inverse mapping. The aggregated variables are not necessarily matched with a single microscopic state. In fact, it is likely that many microscopic states of the agent population yield the same aggregated variables. This issue needs to be addressed in order to produce realistic forecasts. Two methodologies are proposed and evaluated in this work to produce the mapping between the macroscopic state to the microscopic state.

In this work, we provide and evaluate a general framework to use ensemble-based DA on epidemiological ABMs. A compartmental ABM with spatial structure designed for COVID-19 spread is introduced in Section \ref{sec:epi_models}. A general framework of DA and, in particular, ensemble-based DA is presented in Section \ref{sec:da_framework}. Then we discuss a general methodology to apply ensemble-based DA to ABMs (Section \ref{sec:da_abm}). Particular implementations for this coupling for the specific case of our epidemiological ABM are presented in Section \ref{sec:agent_adj}. These are used to assimilate observations with the ensemble Kalman filter. This system is assessed in experiments using synthetic observations as ground truth (Section \ref{sec:synthetic}), and then we use it on real COVID-19 data from Ciudad Autónoma de Buenos Aires (CABA), Argentina (Section \ref{sec:caba}). 

\section{Methods}
\subsection{Epidemiological modeling} \label{sec:epi_models}

To model disease dynamics, the seminal work of \cite{Kermack1927} represents a population divided in compartments. The basic $SIR$ model considers three compartments related to the disease status of the individuals: Susceptible, Infected and Recovered. This has become common practice in epidemiological modeling. Dividing the population in subpopulations, under the assumption of homogeneity in each of them, allows for a small number of variables to summarize the state of the system. These indicate how many individuals are in each compartment. The standard SIR model can be modified and many different compartment configurations can be set up in order to better represent the main characteristics of different diseases. Analysing the flow between compartments provides a general understanding of the disease dynamics. These models only keep track of the macro-state variables --the number of individuals in each compartment-- but they do not model the individuals themselves. Compartmental models are commonly represented by a system of differential equations which can be integrated to get the time evolution of the disease dynamics.

A variety of compartmental models have been used to model COVID-19 which include different traits and complexities. Metapopulation compartmental models may represent populations with age structure  (e.g., \cite{Evensen2020}), geolocalization (e.g., \cite{Arenas2020}), and may include stochastic transmission. The complex interactions between individuals in the population are mostly averaged out in this type of models because of the assumptions of mean field interactions and well-mixed populations. On the other hand, ABMs aim to model people's behavior, represented by agents, explicitly. Each individual of the population is still classified into compartments because each agent will be labeled with its corresponding category. Infections are caused by the interactions between agents, and labels can be changed accordingly. Usually, what is interesting is not the particular outcome of the simulation of an individual but rather the resulting state of the system as a whole. In this case, the total number of agents at each category at a given time gives the aggregated or compartmental representation of the agent-based system. These compartments are not modeled directly with differential equations: the state of each subpopulation emerges from the individual-based level. ABMs also allow the straightforward introduction of relevant features and complexities to the model. For example, waning immunity or health measures such as lockdowns, contact tracing, or vaccination effects can be implemented in a very explicit fashion \citep{Silva2020}.

\subsubsection{Agent-based model details} \label{sec:seihrd}

The ABM we developed was designed to model disease dynamics, in particular for COVID-19 spread. Individuals are characterized mainly by three properties: disease status, house, and neighbourhood. However, the model framework easily allows to include additional properties such as age structure, occupation, or social stratum.

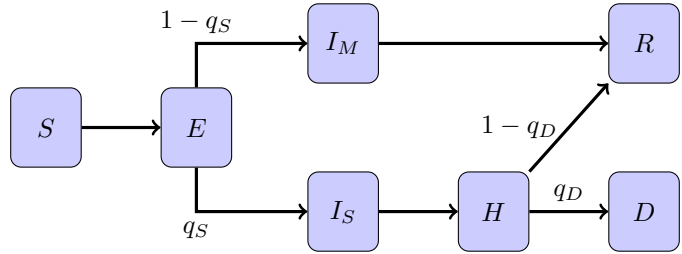
\begin{figure}[b]
    \captionsetup{width=0.5\textwidth}
    \begin{center}
        \tikzstyle{block} = [rectangle, draw, fill=blue!20, 
        text width=4em, text centered, rounded corners, minimum height=3em]
        \tikzset{line/.style={draw, very thick, color=black!100, -latex'}}
        \centering
        \begin{tikzpicture}[node distance = 2cm, auto]
            \tikzstyle{block} = [rectangle, draw, fill=blue!20, 
            text width=2em, text centered, rounded corners, minimum height=3em]
            
            % Place nodes
            \node [block] (S) {$S$};
            \node [block, right of=S] (E) {$E$};
            \node [block, above right = 0.05cm and 1cm of E] (IM) {$I_M$};
            \node [block, below right = 0.05cm and 1cm of E] (IS) {$I_S$};
            \node [block, right of=IS] (H) {$H$};
            \node [block, right of=H] (D) {$D$};
            \node [block] (R) at (IM-|D) {$R$};
            
            % Draw edges
            \draw[line,->] (S) -- (E);
            \draw[line,->] (E) |- node [below] {$q_S$} (IS);
            \draw[line,->] (E) |- node [above] {$1-q_S$} (IM);
            \draw[line,->] (IM) -- (R);
            \draw[line,->] (IS) -- (H);
            \draw[line,->] (H) -- node [above] {$q_D$} (D);
            \draw[line,->] (H) -- node [left = 0.01cm] {$1-q_D$} (R);
        \end{tikzpicture}
    \end{center}
    \caption{Diagram for the epidemiological classes in SEIHRD model: Susceptible ($S$), Exposed ($E$), Mild infected ($I_M$), Severe Infected ($I_S$), Hospitalized ($H$), Recovered ($R$) and Deceased ($D$).}
    \label{diag:seihrd}
\end{figure}

The disease status of each individual is described by 7 categories. Namely, we have the susceptible individuals ($S$) for agents which can get infected, the exposed class ($E$) for those that have been infected but are not yet infectious. The infectious individuals are divided in two groups. The mildly infected class ($I_M$) is intended for individuals that develop  non-hospitalizable form of the disease, including asymptomatic ones, and are expected to recover. The severely infected ($I_S$) are individuals that will require hospitalization. The hospitalized ones ($H$) can recover or die. Finally, we have the recovered ($R$) and the deceased individuals ($D$). We assume that recovered individuals develop immunity for the duration of the simulation, but note that this would be unrealistic for longer-term simulations. Hereinafter, we use these symbols $(S, E, I_M, I_S, R, D)$ to alternatively denote the label of the health status of the individual or the population size of the class. A diagram of the flow between the epidemiological classes is shown in Figure (\ref{diag:seihrd}). 

At every time step, which is considered a day by default, each agent has a number of contacts sampled from a Poisson distribution with parameter $\lambda$. Susceptible agents may become exposed as a result of a contact with an infectious agent: in a contact between agents, if one of them is infectious and the other susceptible, there is a chance that the latter becomes infected too. The time spent in each infected class ($E$, $I_M$, $I_S$, $H$) is considered as a Gamma distribution following \cite{Qin2020}.

The time $\tau_c$ spent by an agent in class $c \in \{ E, I_M, I_S, H\}$ is sampled from a Gamma distribution, i.e. $\tau_c\sim \Gamma(k_c,\theta_c)$ where $k_{c}$ and $\theta_{c}$ are the shape and scale parameters for Gamma distributions. The mean and variance are given by $\mu_{c} = k_{c} \theta_{c}$ and $\sigma^2_{c} = k_{c} \theta_{c}^2$ respectively. When this time $\tau_c$ is spent, the agent will move on to the next epidemiological status. When an agent exits the exposed compartment, it has a chance of $q_S$ of developing a severe sickness and a probability $q_M = 1 - q_S$ of having mild symptoms. In a similar fashion, hospitalized agents have a $q_D$ chance to die and a $q_R = 1 - q_D$ chance to recover.

In addition to the structure given by the health status of the agents, we introduce geographic and demographic information. We consider a city divided into $N_{locations}$ neighbourhoods. Each agent lives in a house in a certain neighbourhood. The house can hold a single agent or it can be shared with others. Contacts are categorized into two cases: domestic and casual contacts. Each of the daily contacts of an agent has a chance $q_C$ of being casual and $1-q_C$ of being domestic. Domestic contacts are between members of the same household and casual contacts are potentially with any other agent. The probability of infection for a domestic contact is $\beta_d$ and  it is $\beta_c$ for casual contacts. In a casual contact, an agent from neighbourhood $j$ will contact another from neighbourhood $i$ with probability $C_{ij}$. So, the connectivity between neighbourhoods is measured by an $N_{locations} \times N_{locations}$ contact matrix $C$. This contact matrix encodes agent mobility between neighbourhoods which in turn is related to work and social activities.  Diagonal terms are expected to be the largest because they quantify the number of contacts within the same neighbourhood. Further structure and classes in the population may be incorporated. For example, age, social profile, occupation or school attendance which can be helpful to represent phenomena such as superspreader events. We have restricted the structure to its minimal expression but keeping house granularity since the main aim of this article is to evaluate the inference technique with widely available data. Because of the effect of higher infection rates and non-viability of distance measures within houses, we considered house granularity was one of the essential individually-based structures to evaluate in the data assimilation system.

\subsubsection{Default parameterization}

Table \ref{table:default_params} summarizes the default parameters used in the experiments. We choose parameters that are representative of COVID-19. However, we do not aim for medical accuracy since the goal of the experiments is to evaluate the methodology in different scenarios. \cite{Guan2020} report a mean incubation period of 4 days, which is consistent with our parameterization. The mean infectious period for mild infections is chosen to be 8 days and values around this are used in other models \citep{Zhao2020, Ivorra2020} and reported in \cite{Byrne2020}. We use 4 days as the mean time between severe illness and hospitalization which is in the range reported in \cite{Faes2020}. The mean time agents spend hospitalized is chosen to be 8.1 days which is compatible with the results in \cite{Vekaria2021}. The default distribution of houses according to the number of inhabitants is given by a vector of probabilities $p_H$ for which the probability of a house to have $i$ members is given by the i-th entry of this vector. Houses with more than 5 individuals are not included in the model, i.e., we assume $p_H$ is of dimension 5. As a default we use $p_H = (0.36, 0.27, 0.16, 0.13, 0.08)$ in both synthetic and realistic experiments, except in the model error experiment (Section \ref{sec:model_error_exp}). These are reference values taken from a demographic survey in CABA, Argentina (specifically Encuesta Anual de Hogares called EAH 2019).

\begin{table}
    \captionsetup{width=0.5\textwidth}
    \begin{center}
        \begin{tabular}{ |c|c|c| }
            \hline
            Parameter & Description & Value \\ 
            \hline
            $\lambda$      & Mean number of daily contacts per agent                 & $0.5$ \\ 
            $\beta_d$      & Infection probability in domestic contact               & $0.8$ \\ 
            $\beta_c$      & Infection probability in casual contact                 & $0.16$ \\ 
            $q_D$          & Death probability for hospitalized individuals          & $0.4$ \\ 
            $q_S$          & Probability of severe symptoms development              & $0.1$ \\ 
            $q_C$          & Probability of a contact to be casual                   & $0.5$ \\ 
            $k_{E}$        & Shape parameter of Gamma distribution for time at $E$   & $1.78$ \\ 
            $\theta_{E}$   & Scale parameter of Gamma distribution for time at $E$   & $2.25$ \\ 
            $k_{I_M}$      & Shape parameter of Gamma distribution for time at $I_M$ & $7.11$ \\ 
            $\theta_{I_M}$ & Scale parameter of Gamma distribution for time at $I_M$ & $1.13$ \\ 
            $k_{I_S}$      & Shape parameter of Gamma distribution for time at $I_S$ & $4.0$ \\ 
            $\theta_{I_S}$ & Scale parameter of Gamma distribution for time at $I_S$ & $1.0$ \\ 
            $k_{H}$        & Shape parameter of Gamma distribution for time at $H$   & $9.0$ \\ 
            $\theta_{H}$   & Scale parameter of Gamma distribution for time at $H$   & $0.9$ \\ 
            \hline
        \end{tabular}
    \end{center}
    \caption{Default parameters for SEIHRD model.}
    \label{table:default_params}
\end{table}

\subsection{Data assimilation framework} \label{sec:da_framework}

In the standard DA framework,  we have noisy partial observations $ \v y_1, ..., \v y_T\in \mathbb R^{N_y}$ of a time evolving process of latent (or hidden) state variables $\v x_0, \v x_1, ..., \v x_T \in \mathbb R^{N_x}$ where the subindices represent discrete times. The initial condition, $\v x_0$, is considered to follow a given distribution $p(\v x_0)$, and is conventionally considered unobserved. The evolution of the state and observations is governed by the state-space model equations for $t = 1, ..., T$:
\begin{align}
    \v x_t &= \mathcal M_{t} (\v x_{t-1}, \gv\eta_t), \label{eq:transition} \\
    \v y_t &= \mathcal H_{t} (\v x_t, \gv\nu_t). \label{eq:observation}
\end{align}
With this nomenclature, the state variables at time $t$ evolve to the following discrete time through a forward model $\mathcal M_t$ which has stochastic components (represented by the random variable $\gv\eta_t$). The relation between data and state is modeled through the observational map $\mathcal H_t$ and $\gv\nu_t$ is a random variable which accounts for the observational error. Equations (\ref{eq:transition}) and (\ref{eq:observation}) represent a hidden Markov model or state-space model. They determine the transition probability $p(\v x_t | \v x_{t-1})$ and the observational likelihood $p(\v y_t | \v x_t)$, respectively \citep{Cappe2006}.

One of the goals of DA is to incorporate the information of observations into the model predictions of the latent state variables. This means that we are interested in the probability of $\v x$ given $ \v y_{1:t}$ (where $\v y_{1:t}\doteq \{\v y_{1},\cdots, \v y_t\}$). In particular, we are interested in the filtering posterior distribution $p(\v x_t | \v y_{1:t})$. This distribution is usually obtained by a two-step iterative procedure: 
\begin{itemize}
    \item Forecasting step:
    $p(\v x_t | \v y_1, ...,  \v y_{t-1}) =
    \int p(\v x_t | \v x_{t-1}) p(\v x_{t-1} | \v y_{1:t-1}) d\v x_{t-1}$ 
    \item Filtering/analysis step:
    $p(\v x_t | \v y_{1:t}) 
    \propto p(\v y_t | \v x_t) p(\v x_t | \v y_{1:t-1})$ 
\end{itemize}

This is a recursive framework: the resulting filtered distribution at time $t-1$ is used to forecast the distribution at time $t$. The forecast is obtained using the forward model $\mathcal M_t$ (forecasting step). In the filtering step, Bayes' rule is used to combine the forecast distribution, which is the prior distribution, with the observation likelihood to update the forecast distribution into the posterior distribution (\cite{Wikle2007}).

\subsubsection{Ensemble-based DA}

The inference approach of DA is probabilistic, so we are interested in the distribution of the state given the observations. If the resulting filtering distribution is Gaussian, then it would be enough to estimate a mean and covariance matrix to represent this distribution. The Kalman filter gives an exact solution when the prior distribution and the observational likelihood function are Gaussian (which in turn results in a Gaussian filtering distribution). This is guaranteed when the operators $\mathcal M_{t}$ and $\mathcal H_{t}$ are linear and the stochastic components $\gv\eta_t$ and $\gv\nu_t$ are additive Gaussian white noise.
In this case, the classical Kalman filter produces a sequence of means $\{\v x^a_t\}_{t=1}^T$ and covariances $\{\v P^a_t\}_{t=1}^T$, such that $p(\v x_t | \v y_{1:t}) \sim \mathcal{N}(\v x^a_t, \v P^a_t)$ \citep{Kalman1960}.

For a non-parametric representation of the distributions, Monte Carlo approaches represent the distributions by a sample. Particle based methods use an ensemble of particles (or ensemble members) to keep track of the forecasting and filtering distribution. These distributions are then represented by an ensemble of states. The general procedure for these methods is to evolve each particle forward using the model to get the ensemble representation of the forecasting distribution and then to transform the states of this ensemble into a sample of the filtering distribution using the information of the observation at that time. This general methodology is specified in Algorithm (\ref{algo:ensemble_forecast_filter}). The procedure used to transform the forecasting ensemble into a filtering ensemble yields different sequential ensemble-based methods. One feature of this framework which will be key for the application to ABMs is that the transition model is basically treated as a black box. This is not the case for the standard Kalman filter for which the linear model is needed explicitly in matrix form for the forecasting to filtering distribution transformation. Two important sample-based families of methods which follow Algorithm (\ref{algo:ensemble_forecast_filter}) stand out: EnKFs and particle filters. If the prediction and observational processes in Eqs. (\ref{eq:transition}) and (\ref{eq:observation}) are weakly nonlinear, it is possible to assume that Gaussianity is preserved through the model's time evolution. Thus, the particles in the filtering step may still assumed to follow Gaussian constraints. This derives in what is known as the EnKF. On the other hand, particle filters do not make any assumptions on the likelihood and prior distributions. They produce a filtered sample by applying Bayes' rule in a fully non-parametric manner \citep{Gordon1993}.

\begin{algorithm}[H]\label{algo:ensemble_forecast_filter}
\SetAlgoLined
  Sample initial particles:
  $\{\v x_{0}^{a(j)} \}_{j=1}^{N_p} \sim p(\v x_0)$\\
  \For{$t=1, ..., T$}{
    \For{$j=1, ..., N_e$}{
        $\v x_{t}^{f(j)} = \mathcal M_t (\v x_{t-1}^{a(j)}, \gv\eta_{t-1})$ 
        using $\mathcal M_t$ from Eq. (\ref{eq:transition}) 
    }
    Transform $\{\v x_{t}^{f(j)} \}_{j=1}^{N_p}$ into $\{\v x_{t}^{a(j)} \}_{j=1}^{N_p}$ using $\v y_t$
  }
  \captionsetup{width=0.5\textwidth}
  \caption{General forecasting-filtering scheme for ensemble DA}
\end{algorithm}

\subsubsection{Parameter estimation through state augmentation}

When the state of the system is partially observed, DA uses the correlation between observed and unobserved variables to propagate the observational information and improve the estimate of variables that are not observed.  With this idea in mind, unknown model parameters can also be interpreted as unobserved variables, and if these are correlated with observed variables, the DA system will calibrate the parameters to values consistent with the observations. To do this, the state is augmented with the parameters, so instead of the state vector $\v x_t$ we use the augmented vector $\tilde{\v x}_t\doteq (\v x_t, \gv \theta_t)$ where $\gv \theta_t$ are the parameters at time $t$. The model operator $\tilde{\mathcal M}$ needs to also account for the evolution of the parameters. A common assumption is to consider that their evolution follows a random walk
$$
\gv \theta_{t+1} = \gv \theta_t + \gv \epsilon_t
$$
where $\gv\epsilon_t$ is Gaussian white noise. Also, $\gv\theta_0$ is considered to be distributed with an a priori distribution based on the range of possible values for $\gv\theta$. A useful feature of this method is that the estimates of $\gv\theta$ can track a parameter that is not constant in time, assuming that the changes are slow (\cite{Ruiz2013}). 

\subsection{Data assimilation in ABMs} \label{sec:da_abm}

The state of an ABM  at a given time $t$ is completely described by the collection of the current values of the attributes for every agent in the ABM. The DA framework previously described cannot be readily applied to this sort of representation because the data attributes which compose agents are not necessarily in a space where DA techniques can be applied. These attributes are computational variables which can be, for example, boolean or categorical data types. A particle filter may be applicable with categorical variables, but we aim at using the EnKF, which only operates in continuous spaces. Even when agent information may be relevant for shaping the spread of the disease, it is likely that the interest is not on the state of each particular individual but rather on an aggregated global information on the population. We perform the DA process in the space of this aggregated information. Actually, these variables are integer values but can be considered continuous when they are large enough. Observations on the system are more likely to be informative on these aggregated variables and not on specific agents.

Considering the population of agents is a discrete set, we refer to a specific agent with the index $k$. The attributes of the $k$-th agent at time $t$ are denoted as $A_{k,t}$. These attributes for each agent define the \textit{micro-state space} $A_{t}=\{A_{k, t}\}_{k=1}^{N_{agents}}$ where $N_{agents}$ is the total number of agents. Given an apriori density of the attributes as a function of the agent classes and the set of model parameters to be calibrated, an ensemble of $N_p$ state members $\{A_t^{(j)}\}_{j=1}^{N_p}$ is drawn representing realizations of the apriori density. Naturally, the micro-state forecast density is constructed evolving the agent-based model starting from the drawn initial micro-states.

The process of DA is conducted in a \textit{macro-state space} $\v x_t \in \mathbb{R}^{N_x}$ composed of aggregated variables and the chosen model parameters. The aggregated (or macroscopic) variables are determined via a map that counts the number of agents in the corresponding class  $\v x_t = \phi_{m\rightarrow M}(A_t)$. In other words, this is a mapping from a micro-structured state ($m$) to a  mean field macro-state ($M$). This mapping is not injective and therefore it is not invertible. This means that we have a model to evolve the micro-state $A_t$ into $A_{t+1}$ and get the resulting $\v x_{t+1}$ by performing the aggregating operations. However, once we correct these macro-state variables with DA, we do not have a unique and explicit model to transform them back into the micro-state variables. We denote this macro to micro map as  $A_{t+1}=\phi_{M\rightarrow m}(\v x_{t+1})$.  We discuss two different approaches to define the macro to micro map in section \ref{sec:agent_adj}. It is important to note that the EnKF and the particle filter allow the model to be treated as a black box so that the filter is unaware of the mapping.

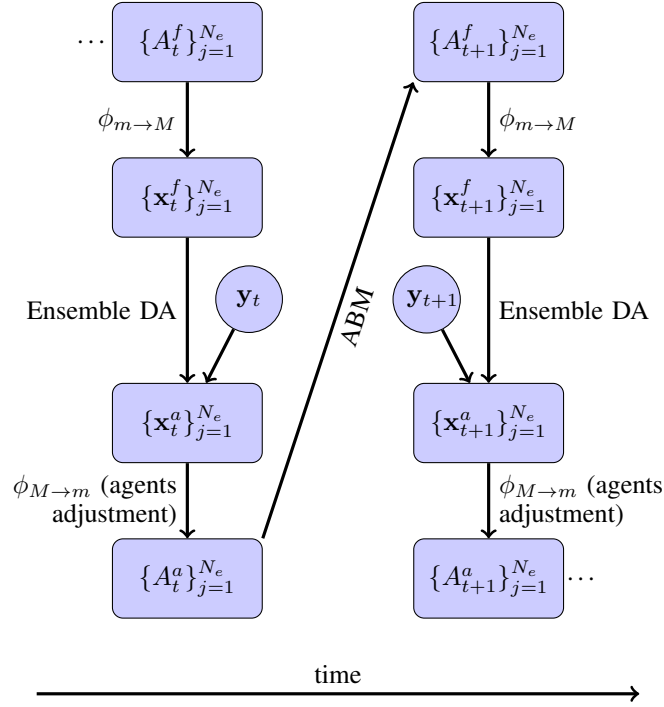
\begin{figure}
\captionsetup{width=0.5\textwidth}
\begin{center}
\tikzset{line/.style={draw, very thick, color=black!100, -latex'}}
\begin{tikzpicture}[node distance = 3cm, auto]
    \tikzstyle{block} = [rectangle, draw, fill=blue!20, 
    text width=5em, text centered, rounded corners, minimum height=3em]
    \tikzstyle{observation} = [circle, draw, fill=blue!20, 
    text width=1.5em, text centered, rounded corners, minimum height=1em]

    % Place nodes
    \node [block] (Af) {$\{A_t^f\}_{j=1}^{N_e}$};
    \node [block, below = 1cm of Af] (xf) {$\{ \v x_t^f\}_{j=1}^{N_e}$};
    \node [observation, below right=0.5cm and -0.5cm of xf] (y) {$\v y_t$};
    \node [block, below of=xf] (xa) {$\{ \v x_t^a\}_{j=1}^{N_e}$};
    \node [block, below = 1cm of xa] (Aa) {$\{A_t^a\}_{j=1}^{N_e}$};

    \node [block, right=2cm of Af] (Af1) {$\{A_{t+1}^f\}_{j=1}^{N_e}$};
    \node [block, below = 1cm of Af1] (xf1) {$\{ \v x_{t+1}^f\}_{j=1}^{N_e}$};
    \node [observation, below left=0.5cm and -0.5cm of xf1] (y1) {$\v y_{t+1}$};
    \node [block, below of=xf1] (xa1) {$\{ \v x_{t+1}^a\}_{j=1}^{N_e}$};
    \node [block, below = 1cm of xa1] (Aa1) {$\{A_{t+1}^a\}_{j=1}^{N_e}$};

    \coordinate[below left = 1cm and 1cm of Aa] (aux);
    \coordinate[below right = 1cm and 1cm of Aa1] (aux1);
    \coordinate[left = 0.5cm of Af] (prev);
    \coordinate[right = 0.5cm of Aa1] (follow);

    % Draw edges
    \draw[line,->] (Af) -- node [left] {$\phi_{m\rightarrow M}$}  (xf);
    \draw[line,->] (xf) -- node [left, text width=2cm, align=right] {Ensemble DA}  (xa);
    \draw[line,->] (xa) -- node [left, text width=3cm, align=right] {$\phi_{M\rightarrow m}$ (agents \\ adjustment)}  (Aa);
    \draw[line,->] (y) -- ($(xa.north) + (0.25cm, 0cm)$);

    \draw[line,->] (Aa.north east) -- node [sloped, below, midway] {ABM} (Af1.south west);

    \draw[line,->] (Af1) -- node [right] {$\phi_{m\rightarrow M}$} (xf1);
    \draw[line,->] (xf1) -- node [right, text width=2cm, align=left] {Ensemble DA} (xa1);
    \draw[line,->] (xa1) -- node [right, text width=3cm, align=left] {$\phi_{M\rightarrow m}$ (agents \\ adjustment)} (Aa1);
    \draw[line,->] (y1) -- ($(xa1.north) - (0.25cm, 0cm)$);

    \draw[line,->] (aux) -- node {time} (aux1);
    \path (prev) -- node[auto=false]{\ldots} (Af);
    \path (Aa1) -- node[auto=false]{\ldots} (follow);

\end{tikzpicture}
\caption{Ensemble based DA for ABMs}
\label{diag:ensemble_DA_ABM}
\end{center}
\end{figure}

At the cycle at time $t$, let us suppose we have an ensemble of micro-states $\{A_t^{f(j)}\}_{j=1}^{N_p}$ representing $N_p$ forecasts at time $t$. Each ensemble member $A_t^{f(j)}$ represents an agent population, where $f$ stands for forecast. We can get a state forecast by applying $\phi$ and get state variables $\{\v x_t^{f(j)}\}_{j=1}^{N_p}$. Using this forecast and the observations $\v y_t$, we can obtain a sample of the posterior distribution, $\{\v x_t^{a(j)}\}_{j=1}^{N_p}$, so-called analysis states, through DA (the $a$ supraindex stands for analysis). The next usual step in DA would be to predict the state at time $t+1$ by evolving $\{\v x_t^{a(j)}\}_{j=1}^{N_p}$ into $\{\v x_{t+1}^{f(j)}\}_{j=1}^{N_p}$. The ABM cannot directly evolve the state variables $\v x_t$ into $\v x_{t+1}$ but rather it can evolve $A_t$ into $A_{t+1}$. Because of this, we need an agent representation $A_t^{a(j)}$ of the analysis macro-state variables $\v x_t^{a(j)}$. This agent set is then used to evolve forward in time and get a forecasted population of agents for time $t+1$, $A_{t+1}^{f(j)}$. In this work, we propose two methods to adjust the forecasted agent population $A_t^{f(j)}$ to be consistent with the analysis state variables $\v x_t^{a(j)}$. With this adjustment, we get the analysis representation of the agent population $A_t^{a(j)}$ which we can now evolve forward with the agent-based model and get the desired forecasted agent population $A_{t+1}^{f(j)}$ completing the forecast-analysis sequence. This methodology is summarized in Figure \ref{diag:ensemble_DA_ABM}.

\subsection{Agents adjustment}\label{sec:agent_adj}

The main issue with the DA cycle is the macro-to-micro map $\phi_{M\rightarrow m}$. For each $j = 1, ..., N_p$ we  have a  mismatch between $\v x_t^{f(j)} = \phi_{m\rightarrow M}(A_t^{f(j)})$ and $\v x_t^{a(j)}$. Our approach to get the filtered ensemble of agent populations $\{A_t^{a(j)}\}_{j=1}^{N_e}$ is to use the forecasted state of agents $\{A^{f(j)}\}_{j=1}^{N_e}$ and change the least possible number of labels to match the analysis macro-state variables $\v x_t^{a(j)}$. This is inspired by the idea that the agents representing the analysis are a correction of the agents representing the forecast. The better the forecast is, the fewer agents have to be changed. However, the inner structure of agents can be very complex, with many other different  attributes aside from the epidemiological status requiring an adjustment. Whether or not it is possible to adjust these attributes realistically will depend on the ABM and on how much of the agent's inner structure is represented by the macro-state variables. 

In order to match the agent state to the filtered state, in the first proposed correction method, the health state classes which lack agents will take the necessary number of agents from the compartments which show an excess. The agents are selected randomly and because of this the method is called \textit{randomized redistribution}. This is achieved through a change of labels and the procedure is repeated independently for every location. Furthermore, not only the labels have to be changed, but other attributes may need to be changed as well. For instance, every agent in either $E$, $I_M$, $I_S$ or $H$ has a time counter which counts the remaining time in the current category and, when it expires, it indicates that the agent has to leave its compartment and enter the next one. These time counters are originally sampled from Gamma distributions, as explained in Section \ref{sec:seihrd}. Then, when an agent is changed in the adjustment to one of these categories its counter has to be reset. Our choice for this is to sample this counter from the current distribution of the counters of agents already in this category. We make this choice in order to be as least intrusive as possible with the agent populations. Although this implementation is particular to this problem (the proposed SEIHRD model), a similar approach can be taken in general as long as we have some prior knowledge on the distribution of values of particular attributes. An advantage of this method is that the amount of agents that change category is the minimum possible.

We implement a second method to adjust the agents which does not necessarily minimize the amount of changes needed but it aims at preserving the history of each individual agent. The choice of the agents which change epidemiological category attempts to select the most likely agents to suffer a change. Changes are made only between adjacent health classes in the SEIHRD progression chain, starting from the latest categories ($R$ and $D$) and ending in the first class ($S$). Also, the selection of the agents to be changed is not random. Considering the time counter of each agent in its category, if a correction needs to be done in the direction of the flow of the diagram, the agents that spent more days in their current category will be transferred to the next. Conversely, if a change is required in the opposite direction, the agents with fewer days in their category are selected to return to the previous compartment. The idea for this criterion is to preserve individual trajectories on the flow of the infection dynamics. For the transitions between susceptible and exposed agents, the time criterion in the susceptible class does not apply. In this case, the criterion to move agents from the susceptible to exposed categories is the number of risky contacts it had (i.e. the number of contacts it had with infectious agents, which do not always result in an infection). In the opposite direction, to move agents from the exposed category to the susceptible we keep the criterion based on the days it spent in the category. Whenever an agent changes its category, the  counter is reset with new values sampled from the corresponding attributes of the population where it has been reassigned. The procedure is applied to each location separately. The method is named \textit{backward cascade redistribution}. The downside for this method is that when we select agents with too few or too many days in their categories to change their status we are cutting off the tails of the distribution of the time spent in each epidemiological class.

\section{Experiments and results}

We conducted a set of experiments using the ABM described in Section \ref{sec:seihrd} coupled with the proposed methodology based on the EnKF. First, we use synthetic observations (Section \ref{sec:synthetic}) to evaluate the overall performance of the system, in which the true state is generated with the ABM with a set of prescribed parameters. In Section \ref{sec:caba}, we use data from the reported cases in CABA, Argentina.

For each neighbourhood, we have 7 macro-state variables which correspond to the variables $(S, E, I_M, I_S, H, R, D)$. So if we consider $N_{locations}$ neighbourhoods, the dimension of the state will be $N_x \times N_{locations}$ without counting the possible parameters which augment the state for parameter estimation. Default parameter values are listed in Table \ref{table:default_params}. If there is no explicit mention of these, it means the value is from the table. 

The EnKF is rather robust to noisy observations, indirect information (e.g., nonlinear observational operator), and partial observations (incomplete state). In the experiments, we assumed the cumulative confirmed cases for each location is observed, which is defined as the sum of $I_M, I_S, H, R, D$ for each neighbourhood. The cumulative number of deaths per location is also observed. The error in the observations is considered to be zero-mean Gaussian noise with variance proportional to the observed quantities themselves. The coefficients of proportionality used are named $\kappa_{C}$ and $\kappa_{D}$, respectively. The possibility of asymptomatic undocumented cases in the cumulative confirmed cases is considered in Section \ref{sec:asymptomatic}.

There are several versions of the EnKF. In this work, because the state space has relatively small dimensions, we use a classic implementation called  EnKF with perturbed observations (\cite{Burgers1998}). Because of sampling errors and unrepresented model errors in the prediction state ensemble, the prediction sample covariance is underestimated, so that  the EnKF ensemble usually has a tendency to collapse, i.e. underestimation feedback between prediction uncertainty and posterior uncertainty. To mitigate this effect, the common methodologies are multiplicative covariance inflation (\cite{Anderson1999, Miyoshi2011}) or additive Gaussian noise in the state variables of the ensemble member updates, i.e. additive inflation. After some preliminary experiments, we found that the intrinsic stochasticity of the SEIHRD ABM results in forecasts with enough spread so that covariance inflation is not required. In other words, the stochasticity in the ABM gives a reasonable representation of the model error. The initial variability of agents populations is given by random sampling at the initialization of agents attributes.

\subsection{Synthetic observations} \label{sec:synthetic}

In these experiments, we produce synthetic observations using the SEIHRD agent-based model and then estimate state variables using the EnKF. Observations are produced using Equations \ref{eq:transition} and \ref{eq:observation}. Since the true state which produced the observations is available, we can examine the technique's performance. In most experiments the model used to simulate the observations is the same as the one used by the EnKF. However, in some experiments, we consider unknown parameters and even some model misspecification.

For these experiments, we use four neighbourhoods, $N_{locations} = 4$. We consider that agents have more contacts within their neighbourhoods. Also, one of the neighborhoods is more concurred by all the agents representing a central neighborhood. This information is encoded in the contact matrix which is consequently set to:
\begin{align*}
    C = 
    \begin{pmatrix}
        0.43 & 0.14 & 0.14 & 0.29 \\
        0.14 & 0.43 & 0.14 & 0.29 \\
        0.14 & 0.14 & 0.43 & 0.29 \\
        0.14 & 0.14 & 0.14 & 0.57 
    \end{pmatrix}.
\end{align*}

\subsubsection{Time varying number of contacts} \label{sec:varying_lambda_exp}

In this experiment, the number of contacts an agent has in a day, which is encoded in $\lambda$, is considered to decrease  linearly in time in the true simulation. This parameter is assumed unknown for the method and is estimated through state augmentation. The simulation uses $3\cdot 10^4$ agents, $100$ EnKF ensemble members, and the observational error coefficients are set to $\kappa_{C} = 1 \cdot 10^{-5} N_{agents}/N_{locations}$ and $\kappa_{D} = 1 \cdot 10^{-6} N_{agents}/N_{locations}$. 

\begin{figure}
    \captionsetup{width=0.5\textwidth}
    \centering
    \includegraphics[width=0.7\textwidth]{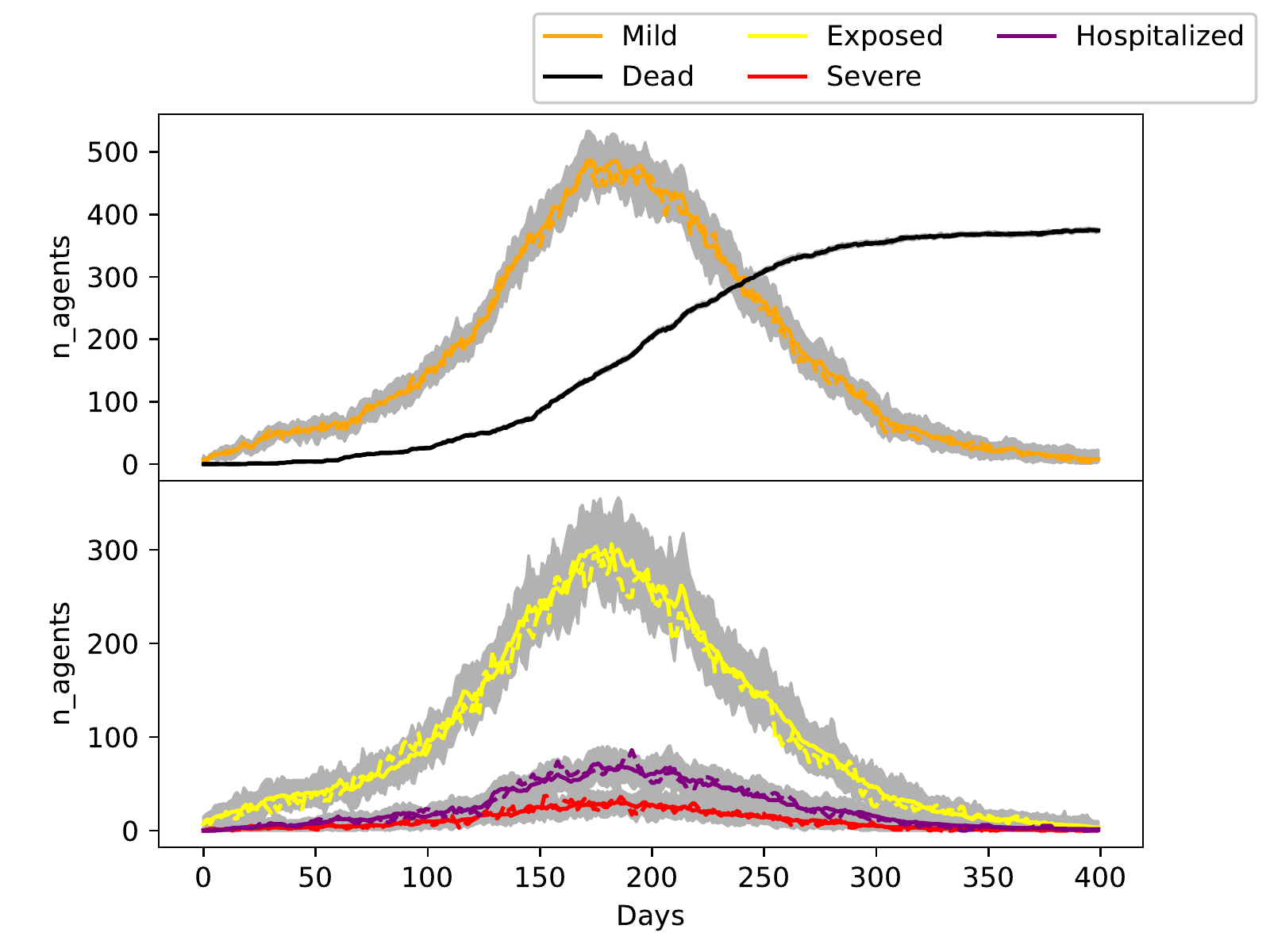}
    \caption{Total number of agents in different epidemiological categories summed over every neighbourhood. Grey lines indicate ensemble members and colored solid lines their mean. True values are indicated with dashed lines.}
    \label{fig:moving_lambda_state_sum}
\end{figure}

\begin{figure}
    \captionsetup{width=0.5\textwidth}
    \centering
    \includegraphics[width=0.7\textwidth]{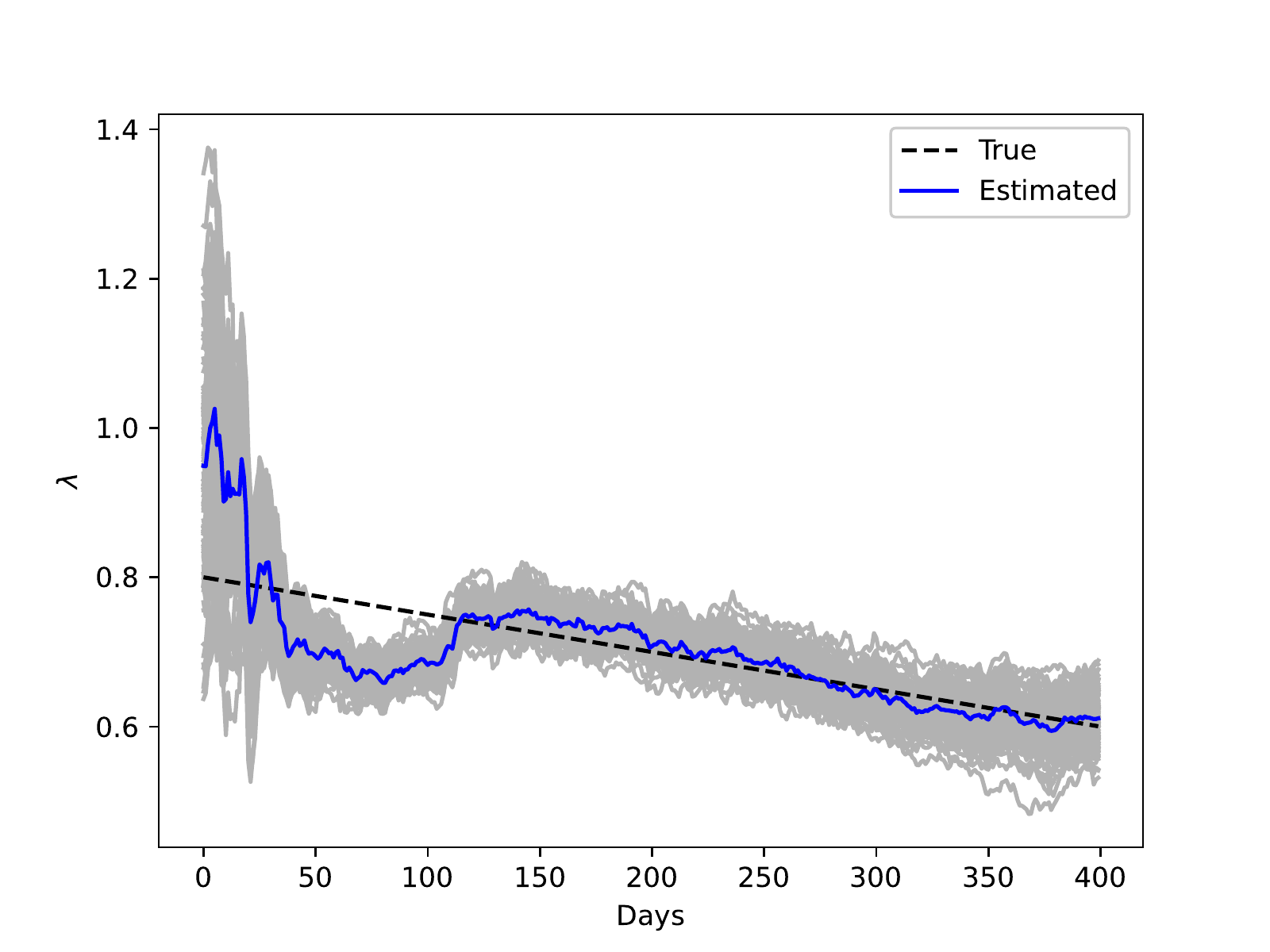}
    \caption{Estimations of $\lambda$ as a function of time shown in solid blue line. Grey lines indicate ensemble members. The true evolution is drawn with a dashed line.}
    \label{fig:moving_lambda_estimates}
\end{figure}

Figure \ref{fig:moving_lambda_state_sum} shows the true trajectories of the state variables and the estimates produced by the EnKF. The filtered ensemble variables are able to correctly estimate the true state variables. The cumulative deaths given by the ensemble members representing the posterior density has very little variance. This is because this variable is directly observed and with a relatively small observational error. In contrast, the other state variables are not observed: they are estimated through the correlations with the observed variables. Figure \ref{fig:moving_lambda_estimates} shows the true value of $\lambda$ over time and the estimates produced by the EnKF with randomized agent redistribution. The estimated values can capture the change of the parameter in time after some initial spin up time. The estimated parameter values are closer to the true value at the epidemic peak. At around 300 days, the variance of the $\lambda$ estimations starts to grow. This is because observations are more informative on $\lambda$ when the number of new infections is higher. After 300 days, the number of active cases is small; correspondingly, the parameter uncertainty increases at those times. The results are shown for the randomized adjustment method but similar results were obtained for the backward cascade method.

\begin{figure}
    \captionsetup{width=0.5\textwidth}
    \centering
    \includegraphics[width=0.7\textwidth]{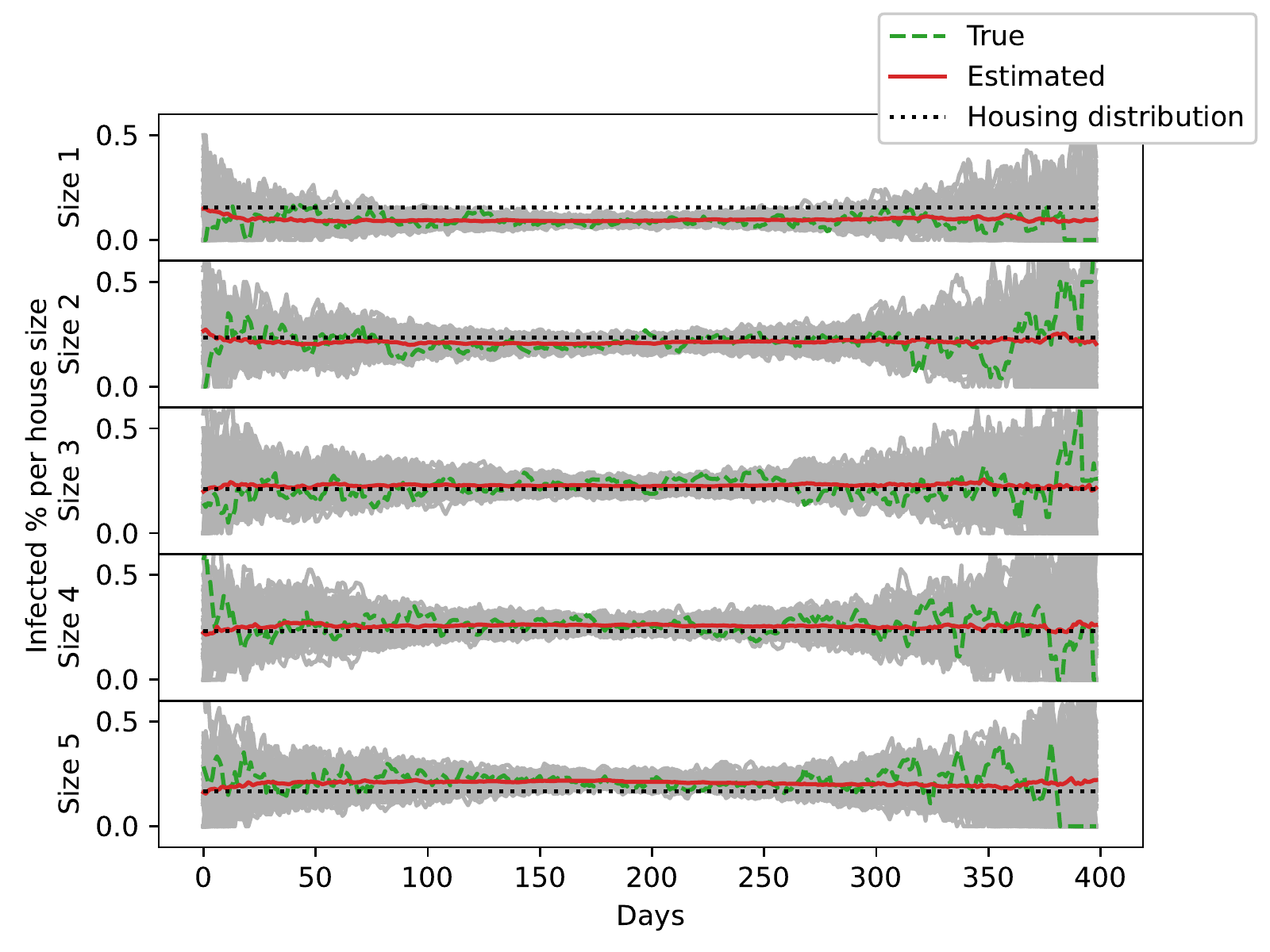}
    \caption{The relative amount of infected agents per house size as a function of time. Ensemble members plotted with grey and their mean with red. The true values are represented with green dashed lines. Dotted black lines indicate the distribution of agents in different house sizes regardless of their disease status.}
    \label{fig:time_histograms}
\end{figure}

The structure given by the distribution of agents in houses is not directly informed on by the observations. However, the model keeps track of how the infections were distributed in different types of houses. The house sizes we considered range from a minimum of a single agent per house to a maximum of five. At each simulation day, we calculated what proportion of the total infected corresponds to each household size. Figure \ref{fig:time_histograms} shows the evolution of the proportion of infections which occurs at each house type. When there are fewer infections, the estimates have higher variance and become more accurate and with less dispersion at the time of the epidemic peak. The proportion of infections in larger houses is greater at the beginning of the epidemic and lower by the end. We can compare the proportion of infected in each house type with the number of agents residing in them regardless of their disease status. In smaller houses, the estimated proportion of infections is less than the proportion of agents residing in that particular house types. The opposite effect occurs with larger houses. This is due to the fact that, because of domestic contacts, the infections spread faster in houses with more members. 

\subsubsection{Assessment of microscale tracking}

We conducted an experiment to evaluate the proportion of agents with matching epidemiological states (between true and estimated agent populations) at different levels of aggregation. The aim of this experiment is to compare the evolution of the agent micro-states between the true agent population that produced the (synthetic) observations and the agent micro-states obtained in the ensemble members of the filter. For this comparison, we count the number of agents with the same epidemiological status, but this count is carried out by considering different groupings for the agents. The first metric (agent\_id) considers agents id per id. The second metric (house\_id) groups agents according to their house id so we count the number of matches house per house. The third metric (household\_type) groups by house size and the fourth (loc\_household\_type) by house size and location. To count the number of matches in the metrics which do not distinguish agents by id we simply count the number of agents in each population that are in the same category.

As in previous experiments, we use $N_{locations}=4$ locations but for each different location $i$, we consider a different $\lambda_i$. Specifically, we set $(\lambda_1, \lambda_2, \lambda_3, \lambda_4) = (1.0, 0.8, 0.9, 0.7)$. We use populations of $5\cdot 10^3$ agents, $100$ EnKF ensemble members, and the observational error coefficients are set to $\kappa_{C} = 1 \cdot 10^{-4} N_{agents}/N_{locations}$ and $\kappa_{D} = 1 \cdot 10^{-5} N_{agents}/N_{locations}$. 

The starting configuration for the agents in the true run and the ensemble members is the same, so the matching proportion at the initial timestep is $100\%$ for every metric. This amounts to have a Delta distribution for every agent population. To evaluate the impact of the assimilation with respect to control simulations, we produce 100 trajectories with the model without any DA. These will yield different results to the true trajectory because of the model's stochasticity, so it is useful reference to compare the metrics.

\begin{figure}
    \captionsetup{width=0.5\textwidth}
    \centering
    \includegraphics[width=0.7\textwidth]{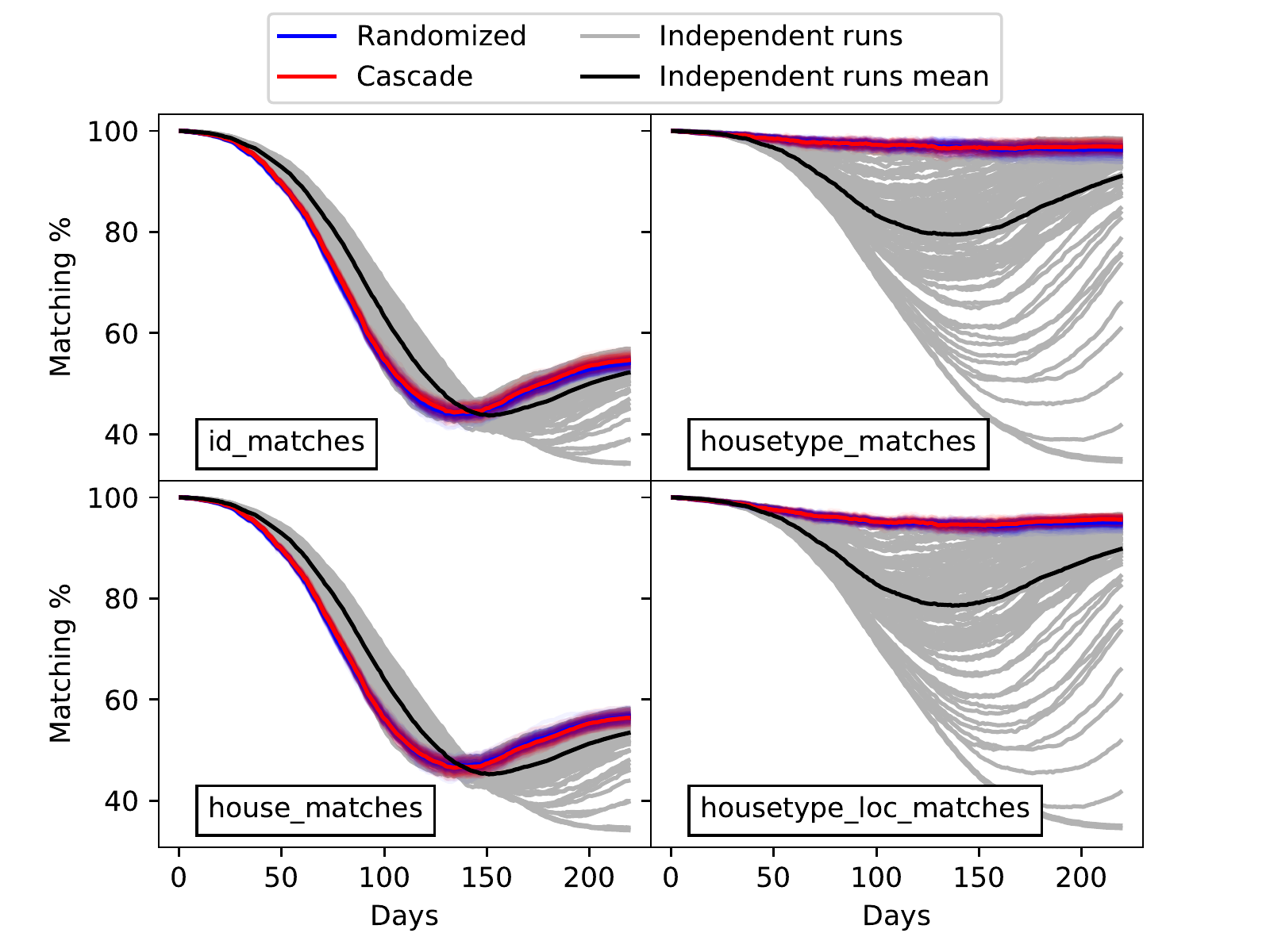}
    \caption{The evolution of the id\_matches, housetype\_matches, house\_matches and housetype\_loc\_matches to the true state are shown from left to right and top to bottom panels. For every panel, the grey lines represent the metrics for the control simulations and their mean is drawn with a black line. Red (blue) transparent lines correspond to EnKF estimation with the randomized (cascade) agent adjustment method. Semitransparent lines represent the ensemble members while their mean is drawn with a solid line of the corresponding color.}
    \label{fig:matches}
\end{figure}

Figure \ref{fig:matches} shows the metrics obtained with the assimilation system for both agent adjustment methods and for the control simulations. Each panel shows one of the different metrics. The randomized and cascade methods are very similar in all the cases. The metrics for control simulations are highly spread out, while  for the ensemble members of the EnKF they are constrained. For the agent\_id and house\_id metrics (these are the metrics which inform on a more microscopic scale) the matching percentage drops, moderately recovers and then stabilizes both for the EnKF and the control simulations. The control simulations are slightly more consistent with the true run that the EnKF runs at the first stage when all metrics decrease. This is likely so because the agent adjustment method by the EnKF has to modify the agent populations in order to match the macro scale while the control simulations maintain the structure of the initial true agent configuration less disrupted.

At the house type scale, the EnKF maintains a high matching proportion (above 90\% for both agent adjustment methods). The control simulations exhibit a large variability. The loc\_household\_type metric yields similar results. Some control simulations mismatch the epidemic peak of the true run. Thus, a decrease of the metrics is expectable for these simulations. The matching proportion recovers at the end of the simulation because the size of the epidemic is similar for the control simulations and the true run, and most agents will be either recovered or susceptible. For the EnKF estimations, a good level of matching is obtained, which is explained because the EnKF is able to track the true state. It is worth noting that observations of the state are classified by location but are not explicitly informative on different house sizes. However, the matching percentage of epidemiological cases at different types of houses is quite high. The proportion of matches when considering agents or houses ids is similar to not using any DA. This is likely because the specific ids of houses and agents have no particular role in the infection dynamics. On the other hand, the house type does have a role in the propagation dynamics and so it is better captured by the EnKF.

\subsubsection{Estimation of undetected infections} \label{sec:asymptomatic}

\begin{figure}[b]
    \captionsetup{width=0.5\textwidth}
    \begin{center}
        \tikzstyle{block} = [rectangle, draw, fill=blue!20, 
        text width=4em, text centered, rounded corners, minimum height=3em]
        \tikzset{line/.style={draw, very thick, color=black!100, -latex'}}
        \centering
        \begin{tikzpicture}[node distance = 2cm, auto]
            \tikzstyle{block} = [rectangle, draw, fill=blue!20, 
            text width=2em, text centered, rounded corners, minimum height=3em]
            
            % Place nodes
            \node [block] (S) {$S$};
            \node [block, right of=S] (E) {$E$};
            \node [block, above right = 0.75cm and 4cm of E] (IA) {$I_A$};
            \node [block, right = 4cm of E] (IM) {$I_M$};
            \node [block, below right = 0.75cm and 4cm of E] (IS) {$I_S$};
            \node [block, right of=IS] (H) {$H$};
            \node [block, right of=H] (D) {$D$};
            \node [block] (R) at (IM-|D) {$R$};
            \node [block] (RA) at (IA-|R) {$R_A$};

            % Draw edges
            \draw[line, ->] (S) -- (E);
            \draw[line,->] (E) |- ++(2.5cm,1cm) |- node [above right = 0cm and 0.4cm, align=center] {$q_A$} (IA);
            \draw[line,->] (E) |- node [above right = 0cm and 0.8cm, align=center] {$1-q_S$} ++(2.5cm,1cm) |- node [above right = 0cm and 0.4cm, align=center] {$1-q_A$} (IM);
            \draw[line,->] (E) |- node [above right = 0cm and 0.8cm, align=center] {$q_S$} (IS);
            \draw [line, ->] (IM) -- (R);
            \draw [line, ->] (IA) -- (RA);
            \draw [line, ->] (IS) -- (H);
            \draw [line, ->] (H) -- (D);
            \draw [line, ->] (H) -- node [left = 0.01cm] {$1-q_D$} (R);
            \draw [line, ->] (H) -- node [below] {$q_D$} (D);
        \end{tikzpicture}
    \end{center}
    \caption{Diagram for SEIHRD model with compartments to account for unreported cases}
    \label{diag:seihrd_unobserved}
\end{figure}
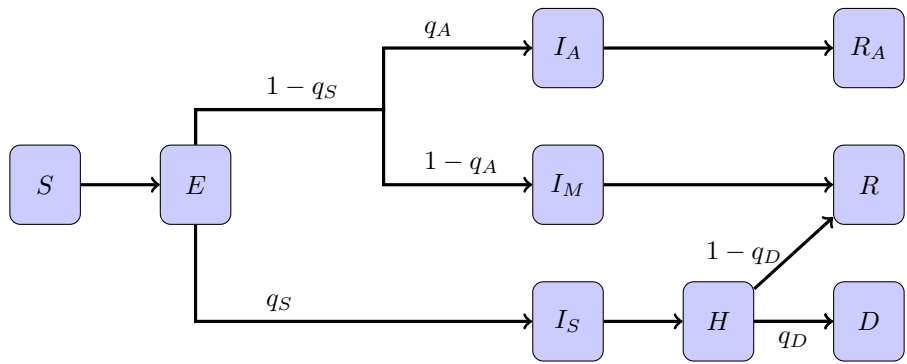

Data on COVID-19 infections is expected to be underreported because of mild or asymptomatic cases. In order to evaluate such scenario, we consider that mild infections have a chance $q_A$ of being asymptomatic, and they are not reported in the data. We use the same setup as in the previous experiment but considering a fixed value $\lambda = 0.8$ and introducing $q_A$ into the augmented state. We change the agent-based model incorporating new epidemiological categories in order to account for this. We incorporate the asymptomatic and asymptomatic recovered epidemiological states. The diagram for the disease development through the epidemiological classes is represented in Figure \ref{diag:seihrd_unobserved} where $I_A$ and $R_A$ stand for infected and recovered asymptomatic, respectively. We note that this scenario could be also represented by assuming unknown coefficients of the observational operator $\mathcal H_t$ and estimating them through state augmentation.

The observed variables are the cumulative confirmed cases (sum of $I_M, I_S, H, R, D$) and deaths ($D$) per neighbourhood. We introduce a new observational variable to the state called global positivity which consists of the proportion of the population which is infected. Asymptomatic cases will be ignored by every observed variable except for the global positivity. To produce observations from this variable in the synthetic experiment,  we simulate data gathered from a randomized COVID-19 testing strategy. At each simulation day, we select a random set of agents and test them. Tests will be positive when the agent is in either $I_M, I_S, I_A,$ or $H$. The tests are not disaggregated by location, so they give an idea of the global circulation of the virus. The error for the observations of the global positivity will be sampling error. We assume 1\% of the population is tested every day. In practice, indirect proxies such as test results of plausible cases (which are not from random samples) could be used to infer global positivity.

For this experiment, we use $3\cdot 10^4$ agents, $100$ EnKF ensemble members, and observational error coefficients $\kappa_{C} = 1 \cdot 10^{-5} N_{agents}/N_{locations}$ and $\kappa_{D} = 1 \cdot 10^{-6} N_{agents}/N_{locations}$. We also consider a fixed value $\lambda = 0.8$. The asymptomatic rate $q_A$ (which here also accounts for undocumented cases) is estimated by augmenting  the state with this parameter. The rest of the configuration is similar to the experiment in Section \ref{sec:varying_lambda_exp}. The asymptomatic rate is expected to be estimated with the EnKF via correlations with the global positivity.

\begin{figure}
    \captionsetup{width=0.5\textwidth}
    \centering
    \includegraphics[width=0.7\textwidth]{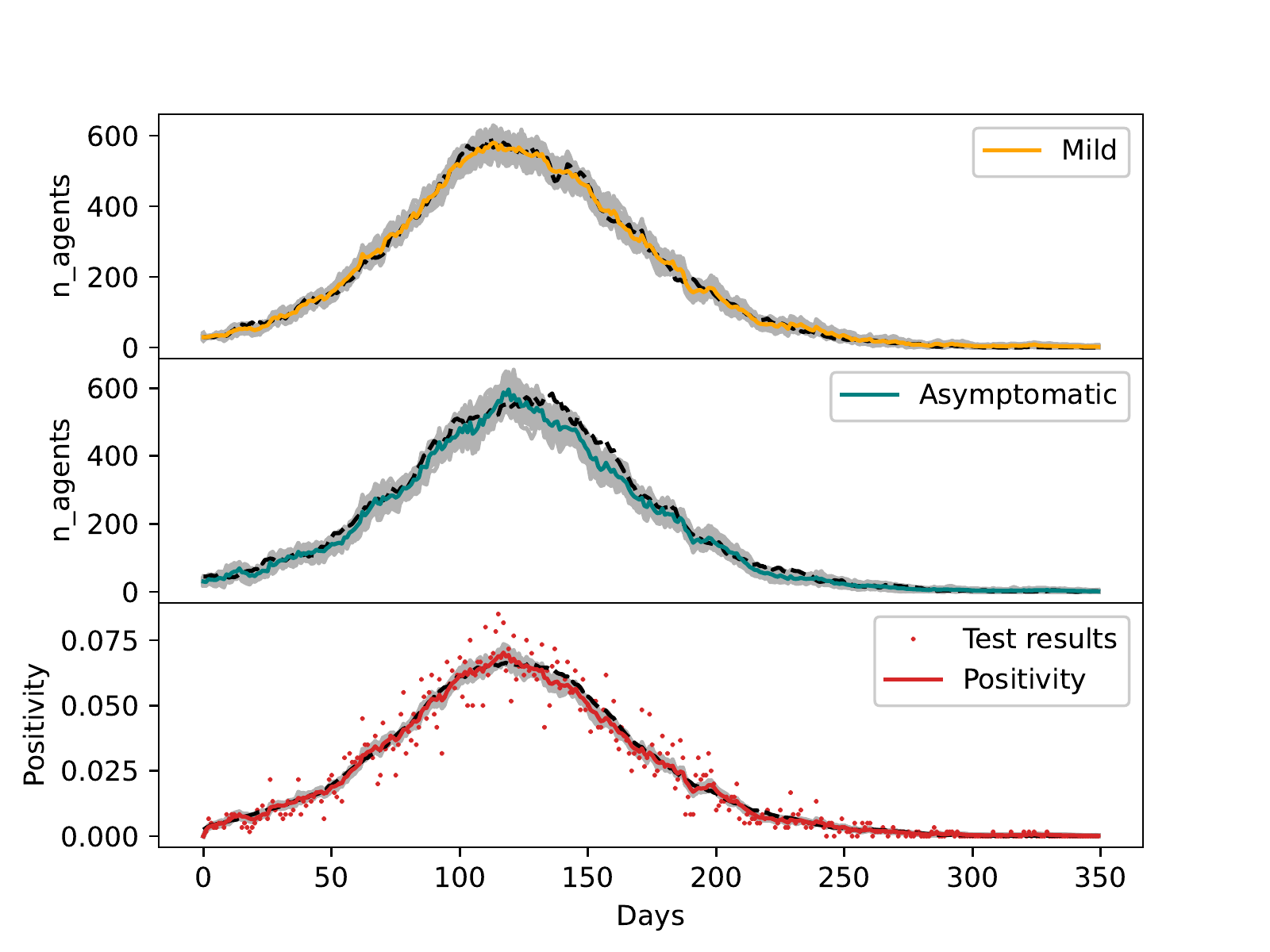}
    \caption{
    Top panel: evolution of reported active mild cases. Orange solid line is the ensemble mean.
    Middle panel: Unreported active cases. Teal solid line is the ensemble mean.
    Bottom panel: Positivity (percentage of agents that would test positive). Red solid line is the ensemble mean. Dots correspond to the testing data generated with the true run.
    For every panel, grey lines indicate ensemble members and dashed lines the true value of the corresponding variable.
    }
    \label{fig:unobserved}
\end{figure}

\begin{figure}
    \captionsetup{width=0.5\textwidth}
    \centering
    \includegraphics[width=0.7\textwidth]{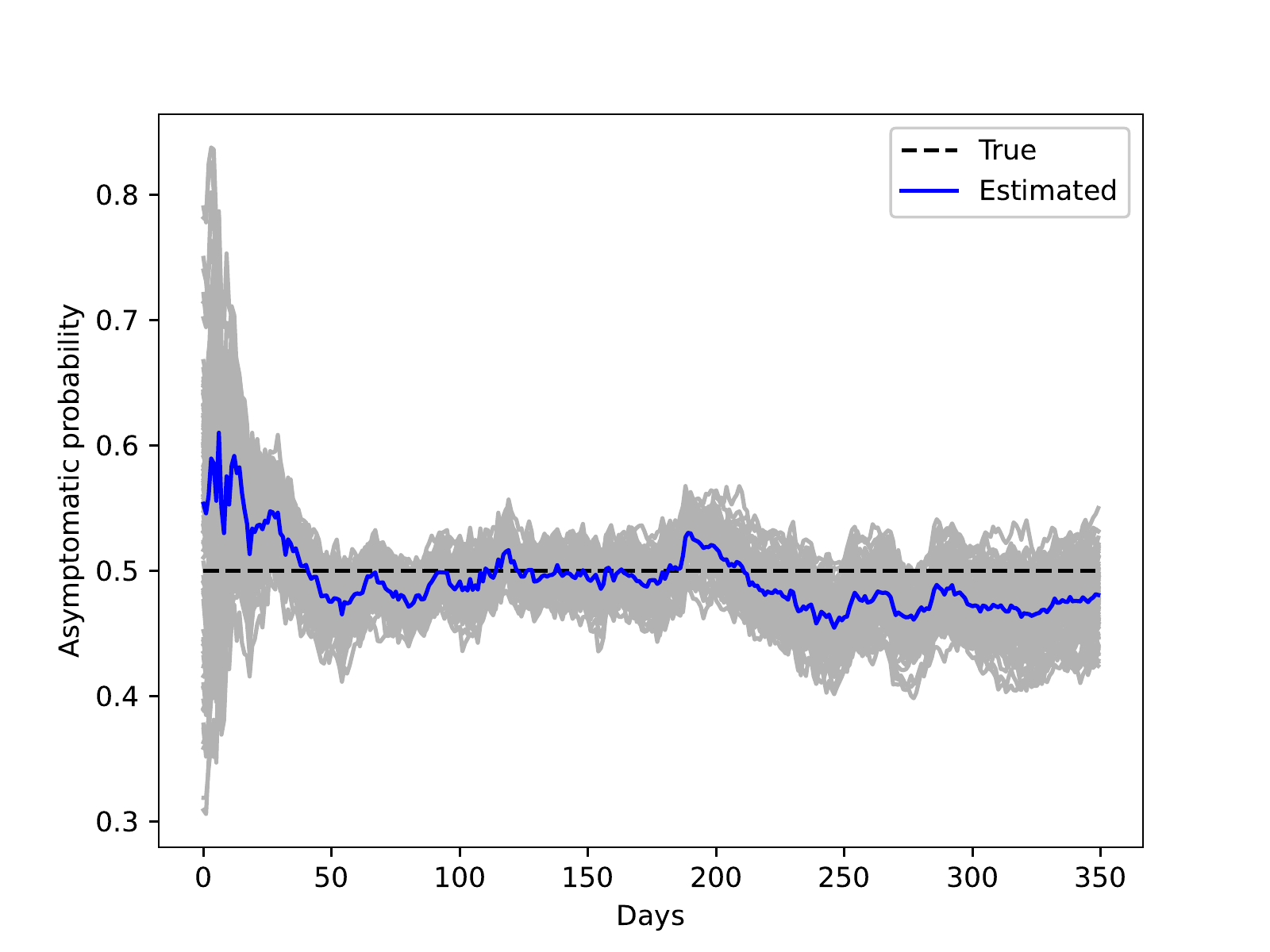}
    \caption{Estimations of $q_A$ as a function of time are shown with solid blue lines, ensemble members with grey lines, and the true parameter value with a black dashed line.}
    \label{fig:asymptomatic_prob}
\end{figure}

Figure \ref{fig:unobserved} shows the evolution of the number of mild infected agents, the asymptomatic, and the global positivity alongside tests' results. The overall behaviour of the asymptomatic agents is rather well captured by the EnKF but, as expected, the estimations are more accurate for the mildly infected than for the asymptomatic cases. This is because the asymptomatic are only informed on by test results while the symptomatic cases are also informed on by the cumulative confirmed cases. The mismatch between the asymptomatic with their true value is correlated with the mismatch between the true and estimated global positivity. This suggests that the more we know about the general circulation of the virus, the better we can infer the underreported cases. Figure \ref{fig:asymptomatic_prob} shows the estimation of the asymptomatic probability which gives noisy estimates around the true value of $q_A$ which is $0.5$. When the positivity is low at the start and end of the epidemic, the estimations of $q_A$ do not synchronize well with the true value and the uncertainty grows, i.e. the ensemble shows more spread. Because the global positivity is correlated with the asymptomatic infected compartment, $I_A$, the system can use these correlations to give an estimate of $q_A$. 

\subsubsection{Model error experiment} \label{sec:model_error_exp}

In order to evaluate whether the inference system is robust enough to account for a misspecification in the agent model setup, we conducted an experiment in which the observations and the EnKF predictions are obtained through different agent-based model configurations. To generate the synthetic observations, the number of daily contacts is sampled from a geometric distribution with parameter $p=0.5$. On the other hand, the model in the assimilation uses a Poisson distribution as in previous experiments. Additionally, we use different household type distributions: the true simulation is conducted with $p_H = (0.33, 0.27, 0.2, 0.13, 0.07)$ for which we have more households with few agents and less households with more agents, which is an expectable situation in real life (close to those used in the rest of the experiments which represents the CABA house distribution). For the EnKF runs we use this distribution but we also repeat the experiment with $p_H = (0.2, 0.2, 0.2, 0.2, 0.2)$ (we call this uniform housing) and $p_H = (0.07, 0.13, 0.2, 0.27, 0.33)$ for which we have more larger households and less households with few people (we call this distribution unbalanced housing). Different configurations for household distributions will yield different propagation dynamics due to variations in contact structures \citep{Grossmann2021}.

The parameter $\lambda$ is assumed unknown and we estimate it through state augmentation. We use $3\cdot 10^4$ agents, $100$ EnKF ensemble members, and observational error coefficients $\kappa_{C} = 1 \cdot 10^{-5} N_{agents}/N_{locations}$ and $\kappa_{D} = 1 \cdot 10^{-6} N_{agents}/N_{locations}$.

\begin{figure}
    \captionsetup{width=0.5\textwidth}
    \centering
    \includegraphics[width=0.7\textwidth]{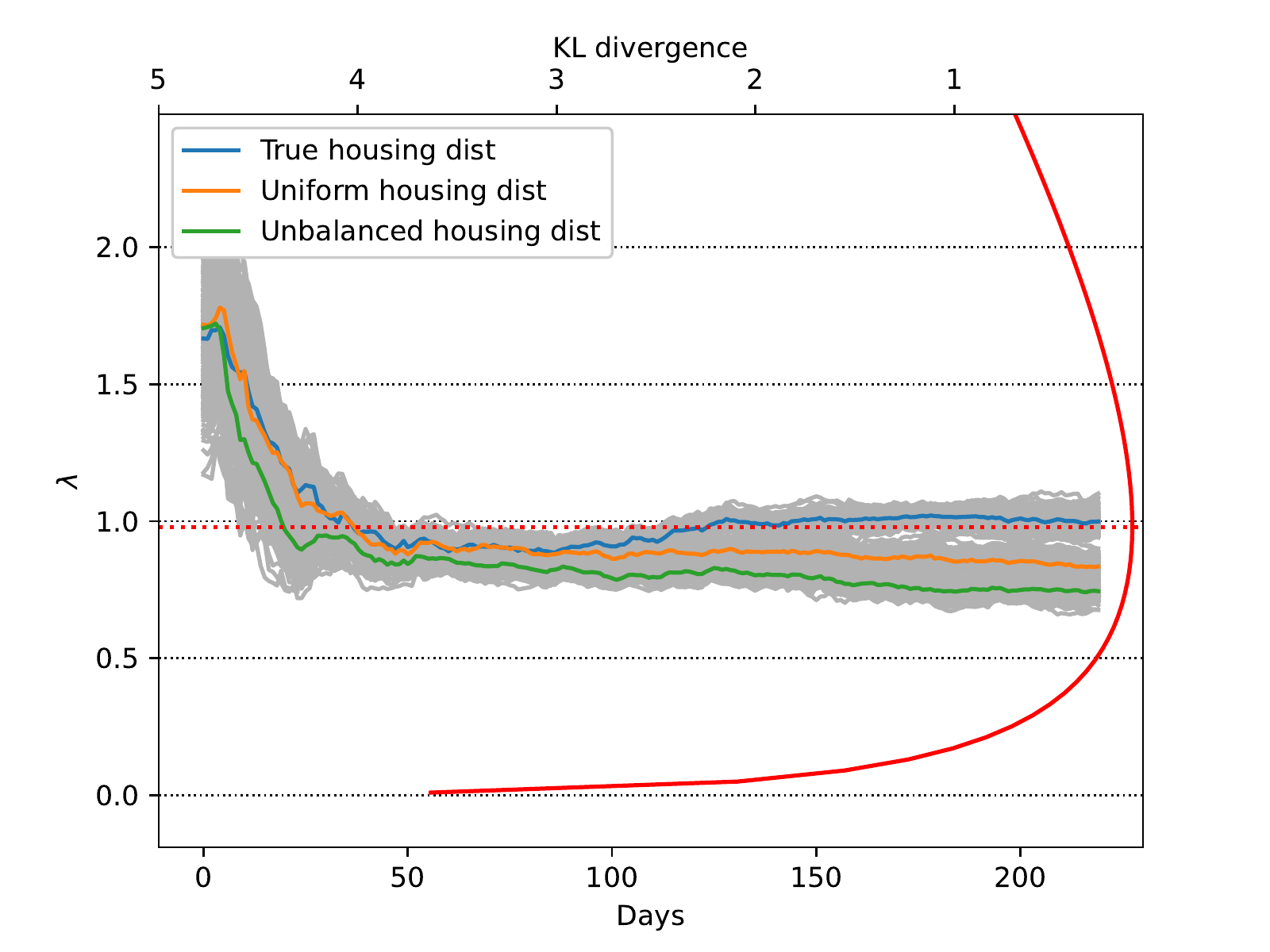}
    \caption{Evolution of the $\lambda$ estimates means for three different housing scenarios shown with solid colored lines. Ensemble member estimates represented with grey lines. The solid red line indicates the KL divergence between a Poisson distribution and the true geometric distribution used. The dotted red line indicates the minimum of the KL curve.}
    \label{fig:KL}
\end{figure}

Figure \ref{fig:KL} exhibits the $\lambda$ estimates as a function of time for each of the three different housing scenarios. In order to compare how the estimated distribution performs, we compute the Kullback-Leibler divergence between a Poisson with and the true geometric distribution as a function of the Poisson parameter $\lambda$. This encodes how much information is lost when we use the estimated Poisson distribution instead of the geometric. Figure \ref{fig:KL} also shows the Kullback-Leibler divergence alongside the $\lambda$ estimates. These are in a region of low KL divergence with respect to the true distribution. This means that the system self-calibrates using a Poisson distribution which causes a model behaviour which resembles model's characteristics when a geometric distribution is used. The best estimate in terms of KL divergence is for the experiment which uses the true housing distribution, which is expectable. The uniform housing distribution has more agents living in larger households and the unbalanced even more. This will produce an effect of faster spread of the disease in the model. The fact that the $\lambda$ estimates are smaller in these cases is because the system is self-calibrating by estimating $\lambda$. A smaller value for $\lambda$  compensates for the misspecification of the housing distribution. Comparatively, the disease dynamics propagate more by intra-house contacts than by external contacts in the latter. 

\subsection{Ciudad Autónoma de Buenos Aires data} \label{sec:caba}

To evaluate the system with realistic observations, we use the data on COVID-19 cases for CABA, Argentina. The data is provided by the Ministry of Health and published in \url{https://data.buenosaires.gob.ar} with daily updates. The distribution of houses and population is taken from census data also publicly available through the same website. CABA is divided into 15 neighbourhoods (or communes) and epidemiological data is disaggregated according to these. We use the number of cumulative confirmed cases and deaths. The observational error variance is considered to be proportional to the observed variables. We estimate the parameter $\lambda$ (which is assumed to be the same for every neighbourhood) through state augmentation. To construct the contact matrix we considered that half of the casual contacts are within each commune and that the other half of the casual contacts is distributed among other communes proportionally to their population density. The number of agents used is $3\cdot 10^5$, and CABA has a population of around $3\cdot 10^6$, so we scale down the data by $10$. We use $400$ EnKF ensemble members. The observational error coefficients are $\kappa_{C} = 5 \cdot 10^{-6} N_{agents}/N_{locations}$ and $\kappa_{D} = 5 \cdot 10^{-7} N_{agents}/N_{locations}$.

\begin{figure}
    \captionsetup{width=0.5\textwidth}
    \centering
    \includegraphics[width=0.7\textwidth]{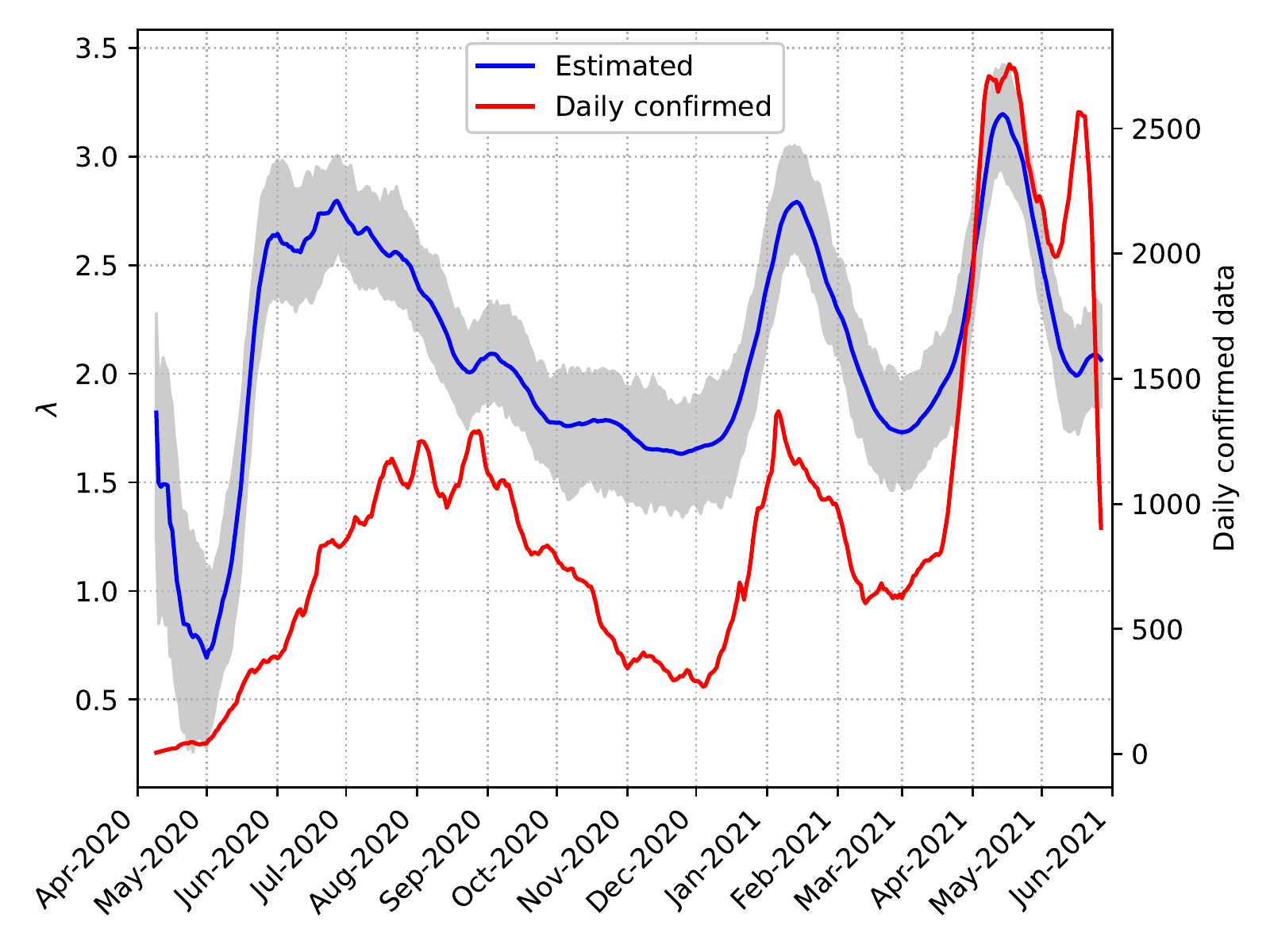}
    \caption{Evolution of $\lambda$ estimations shown in solid blue line and ensemble members with gray lines. Red line corresponds to the 7-day rolling average of daily confirmed cases.}
    \label{fig:caba_r0}
\end{figure}

Figure \ref{fig:caba_r0} shows the estimated $\lambda$. This parameter measures how many contacts, on average, an agent will have on a given day. We can expect this to be correlated to the number of confirmed daily cases, so we plot the data of the daily reported cases alongside the $\lambda$ estimations, and indeed we can see that they follow a similar trend. This happens because  $\beta_d$ and $\beta_c$ values remain constant throughout the simulation, so changes in the number of new cases are determined by changes in the value of $\lambda$ with the corresponding time lag due to the incubation period. In the figure we show a 7-day rolling average of the daily cases but the experiment was performed on the original data without this processing.

The evolution of the estimated contact rate in Figure \ref{fig:caba_r0} is coherent with the three epidemic waves suffered in Argentina. There was a first period of a strong lockdown between April and May, which was then relaxed during austral winter giving place to the first wave. The second wave occurred not surprisingly in January-February which is summer holidays time in Argentina. The strongest wave is the third one, but it started fading because of, among other factors, massive vaccinations starting in May 2021.

It is important to note that of the many time-dependent processes that may affect the development of the disease spread, we only estimate $\lambda$ while keeping other parameters fixed. This means that the calibration of the model to fit the data acts through $\lambda$, even though what drove the effect on the data may have another cause. For example, a decrease in cases because of the use of face-masks should be accounted for by a change in the casual contagion probability $\beta_c$, but since we are keeping this parameter fixed, this change will be captured by $\lambda$. Although this may be inaccurate, we found that trying to estimate parameters with similar effects on data leads to overparameterization and lack of identifiability. 

\begin{figure}
    \captionsetup{width=0.5\textwidth}
    \centering
    \includegraphics[width=0.7\textwidth]{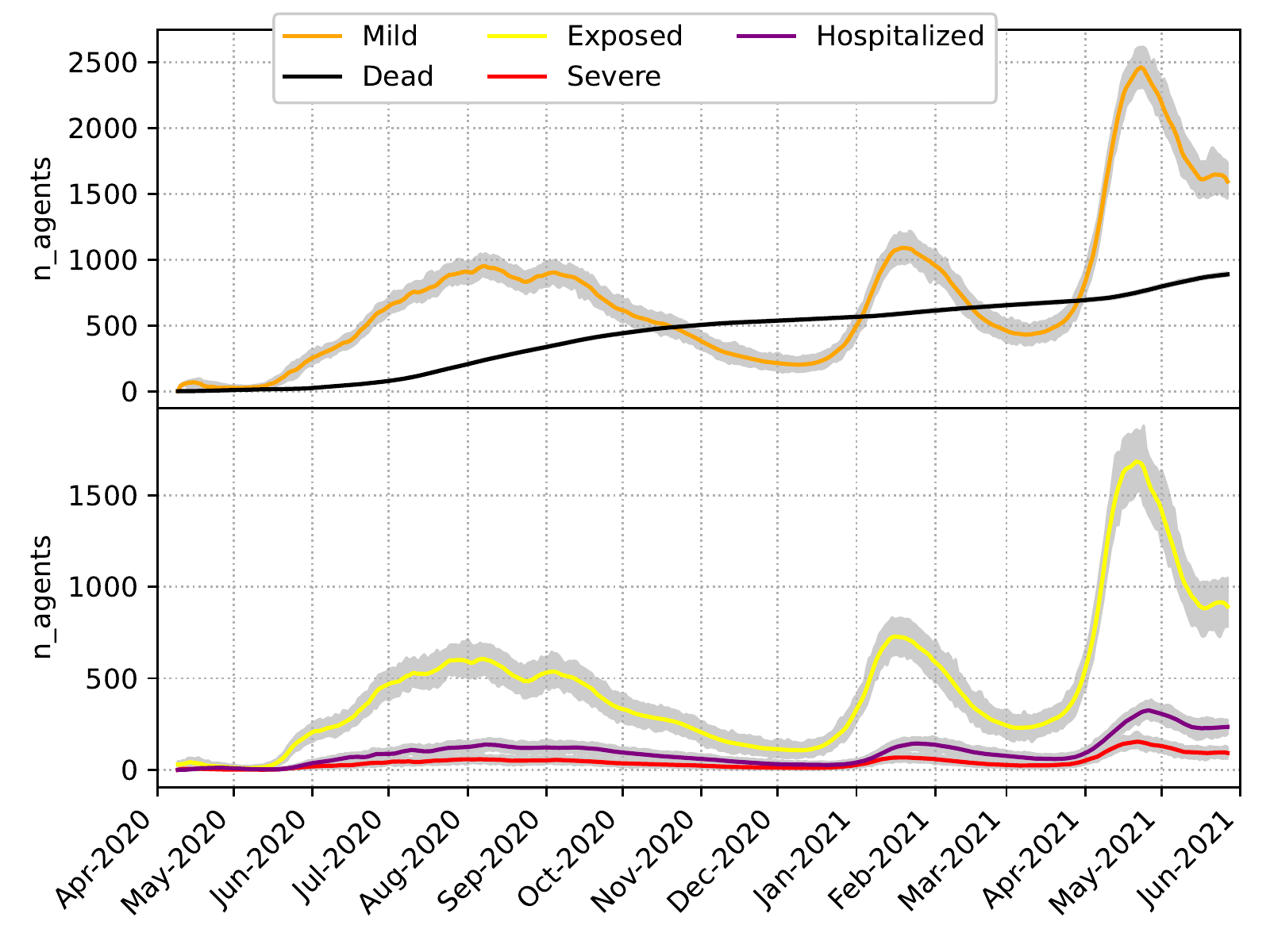}
    \caption{Total number of agents for different epidemiological categories. Grey lines indicate ensemble members and colored solid lines their mean.}
    \label{fig:caba_state_sum}
\end{figure}

Figure \ref{fig:caba_state_sum} shows some of the aggregated state variables of the system summed over every neighbourhood. The cumulative deaths have comparatively less variance than the other variables. Like in the experiments with synthetic observations, this is because deaths are directly observed, and the other variables are only observed through the cumulative cases, which is a sum of several epidemiological states.

\begin{figure}
    \captionsetup{width=0.5\textwidth}
    \centering
    \includegraphics[width=0.7\textwidth]{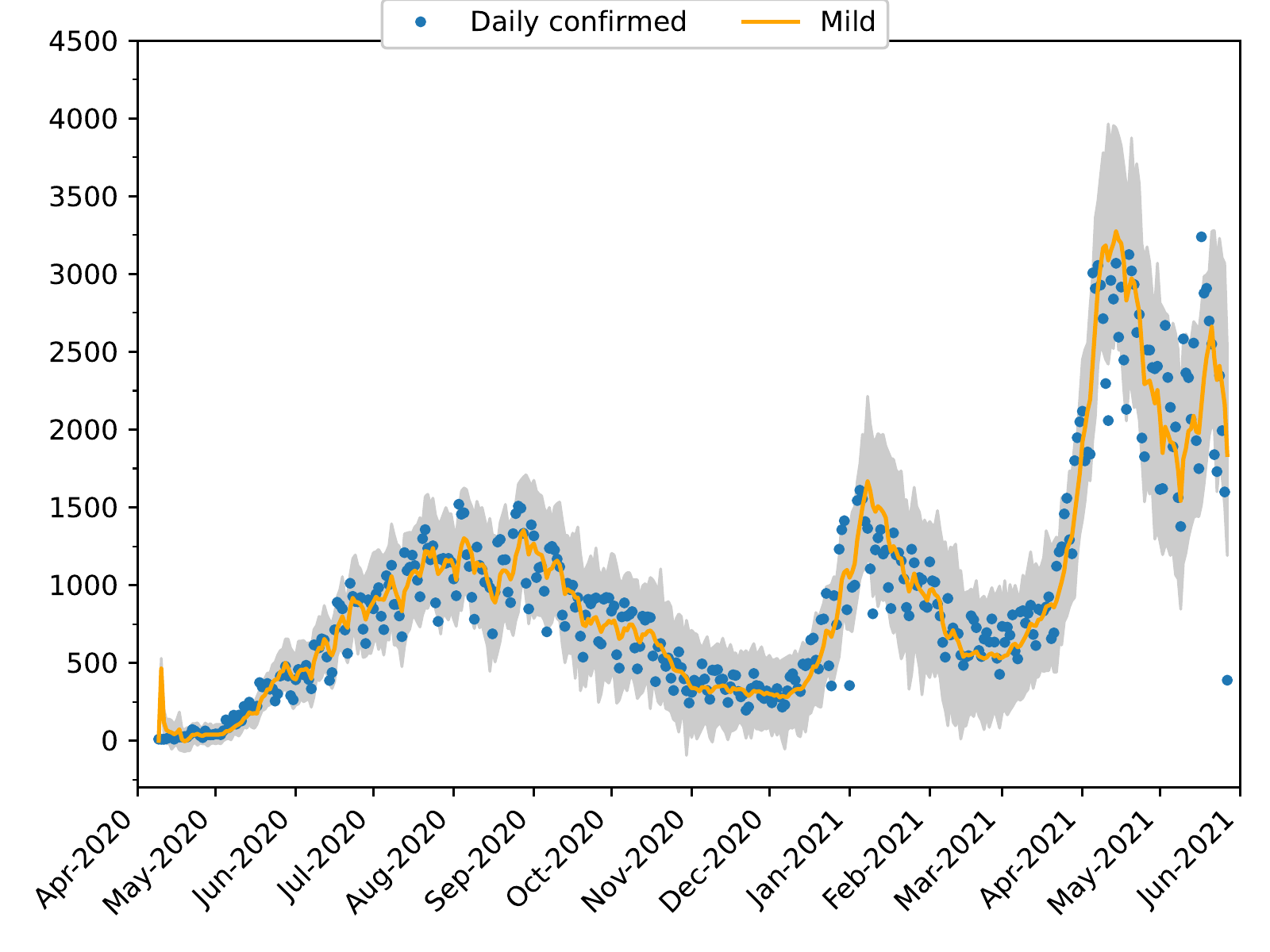}
    \caption{Estimated daily new cases for the whole city are shown with orange lines and the corresponding ensemble with grey. Blue dots correspond to the daily confirmed cases.}
    \label{fig:aggregated_incidence}
\end{figure}

\begin{figure}
    \captionsetup{width=0.5\textwidth}
    \centering
    \includegraphics[width=0.7\textwidth]{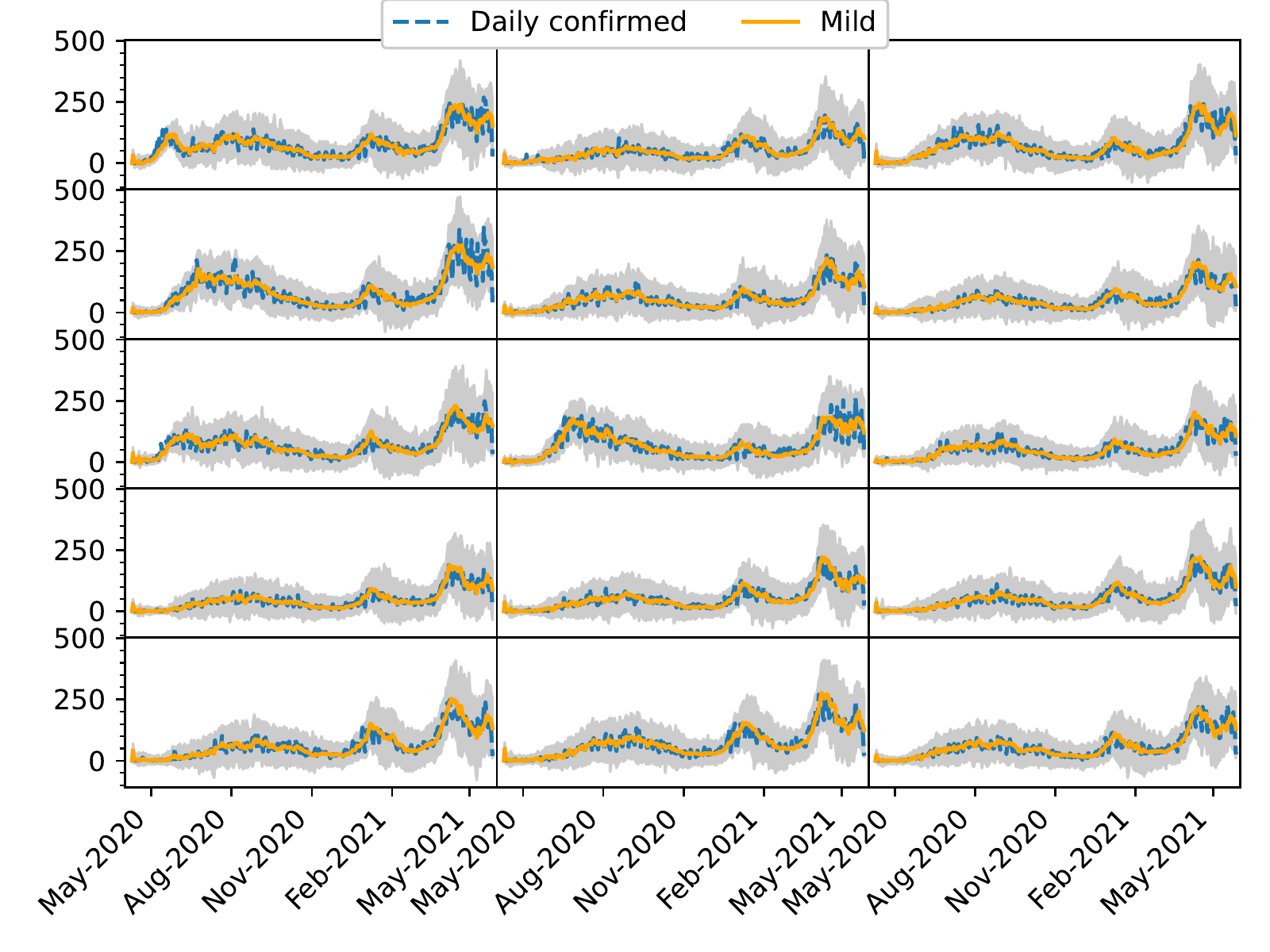}
    \caption{Each plot represents one of the 15 CABA communes. Estimated daily new cases are plotted with orange lines and the corresponding ensemble with grey lines. Blue dots correspond to the daily confirmed cases.}
    \label{fig:neighbourhood_incidence}
\end{figure}

Figure \ref{fig:aggregated_incidence} shows the estimated number of daily new cases as a function of data of confirmed new cases. The data are affected by underreporting on weekends, but the estimations tend to smooth out that effect. Figure \ref{fig:neighbourhood_incidence} shows the same metrics but disaggregated by neighbourhood and we can see that although all the neighbourhoods show similar trends, some have particularities in the shape of the epidemic peaks. These are rather closely captured by the EnKF estimations.

\section{Discussion}

In this work, we propose a general framework for the application of DA techniques to ABMs. The method is designed for cases in which the only observed variables are aggregated information of the agent population state. We applied it to an epidemiological ABM in multiple scenarios coupled with an implementation of the EnKF. The applicability of the method depends on the possibility to adapt the forecasted agent populations to match the macroscopic state variables. In our case, we developed two different methodologies to do this, referred to as randomized redistribution and backward cascade redistribution. They both yielded similar results. We found that it is possible to track the macroscopic state variables even when the system is driven by microscopic dynamics. Also, we show that the state augmentation technique can be used to calibrate the model in a sequential manner, tracking the dynamics of a parameter which changes slowly over time. The methodology also responded well to data produced by the daily monitoring of the COVID-19 pandemic in CABA. In particular, the model calibration adapts the parameterization to fit the changing behaviour of the spreading of the disease.

The experiments in this work show that the EnKF is a robust Gaussian technique for epidemiological ABM data assimilation. However, many of the challenges posed by DA are inherited by the framework we present: for example, specification of model and observational errors or dealing with non Gaussianities. Alternatively to the EnKF, there is a vast variety of DA methods which could possibly be helpful to do inference alongside ABMs. Methods which do not make sequential updates but jointly assimilate data over a time window (for example, ESMDA or pMCMC) could prove useful to circumvent the need of adapting the agent populations to the macroscopic state. Among sequential methods, particle filters which rely on resampling in order to transform the forecast into the filtered ensemble, such as the bootstrap particle filter, could be interesting to apply. The fact that the filtered ensemble is just a resample of the forecast particles means that there would be no need to readjust the agent populations.

The assimilation system shows sensitivity to the distribution of the number of households in the experiments. This is a bottom-up effect produced by the agent-based model which cannot be readily modeled with compartmental models. The coupling of the EnKF with the agent-based model may be useful, in future work, for the modelling of disease propagation in cities with vulnerable neighborhoods of developing countries, in which the proportion of households with a large number of members is much higher.

The computational cost of the methodology can quickly escalate since each ensemble member consists of a whole agent population. However, there is much room for improvement because several tasks may be paralellized (at the agent population level and also at the ensemble level). The computational cost of DA techniques themselves is usually associated with matrix inversions in high dimensions. However, that was not the case because we applied DA to the aggregated state variables, which constitute a relatively small sized state space.

In this work, all the observations are aggregated state variables, but the technique can potentially be applied for inference on the agents using data which is gathered at that micro scale. The fact that more individual-based data is collected through mobile devices is making this sort of data more accessible every day, for example, from digital contact tracing apps \citep{Munzert2021,Lopez2021}. It is expectable that the technique proposed in this work, based on the EnKF, may be able to assimilate these micro-state observed variables.

\clearpage

% \section{Bibliography}

% \printbibliography[heading=none]
\bibliography{bibliography}

\end{document}